\documentclass[aps,tightenlines,nofootinbib,11pt]{revtex4}
\newcommand{\PRE}[1]{{#1}} % Use if preprint style
\usepackage{graphicx}
\usepackage{type1cm}
\usepackage{eso-pic}
\usepackage{color}
\usepackage{subfigure}

\usepackage{hyperref}      % hypertext links %%ARXIV
\usepackage{bm}
\usepackage{epsfig}
\def\tz{\widetilde Z}
\def\ta{\tilde a}
\def\tg{\tilde g}

\def\beq{\begin{eqnarray}}
\def\eeq{\end{eqnarray}}
\def\bea{\begin{eqnarray}}
\def\eea{\end{eqnarray}}

\newcommand{\mweak}{m_W}

\newcommand{\sigmath}{\sigma_{\text{th}}}

\newcommand{\OmegaDM}{\Omega_{\text{DM}}}

\newcommand{\gev}{\text{GeV}}
\newcommand{\tev}{\text{TeV}}

\newcommand{\cm}{\text{cm}}

\newcommand{\g}{\text{g}}

\newcommand{\eg}{{\em e.g.}}

\newcommand{\eqref}[1]{Eq.~(\ref{#1})}

\newcommand{\secref}[1]{Sec.~\ref{sec:#1}}

\newcommand{\figref}[1]{Fig.~\ref{fig:#1}}

\newcommand{\Figref}[1]{Figure~\ref{fig:#1}}

\newcommand{\gsim}{\lower.7ex\hbox{$\;\stackrel{\textstyle>}{\sim}\;$}}
\newcommand{\lsim}{\lower.7ex\hbox{$\;\stackrel{\textstyle<}{\sim}\;$}}
\newcommand{\met}{\rlap{\,\,/}E_T}

\newcommand{\Omegachi}{\Omega_{\chi}}

\newcommand{\mhu}{m_{H_u}}

\def\beqn{\begin{eqnarray}} 
\def\eeqn{\end{eqnarray}} 
\def\be{\begin{equation}}
\def\ee{\end{equation}}

\def\c{\hspace{-5pt}}

\def\n{\chi_1^0}
\def\c{\chi_1^\pm}
\newcommand{\bi}{\begin{itemize}}
\newcommand{\ei}{\end{itemize}}

\newcommand{\ax}{\alpha_\chi}
\newcommand{\mphi}{m_\phi}
\newcommand{\mx}{m_\chi}

\begin{document}

%\renewcommand{\thefootnote}{\fnsymbol{footnote}}
%\setcounter{footnote}{0}

%\preprint{number}

\title{ \PRE{\vspace*{0.15in}} {\Large Dark Matter in the Coming Decade: \\
  Complementary Paths to Discovery and Beyond} \PRE{\vspace*{0.15in}} }

%\author{CF4}
%\affiliation{
%\PRE{\vspace*{.2in}}
%}

%\author{Snowmass 2013 Cosmic Frontier Working Group 4: Dark Matter
%  Complementarity\footnote{Suggestions and corrections are most
%    welcome and should be directed to one or more of the following
%    contributors: Jim Buckley, Jonathan Feng, Manoj Kaplinghat,
%    Konstantin Matchev, Dan McKinsey, and Tim Tait.}}

\author{
    {\bf Sebastian Arrenberg}, University of Z\"urich; 
    {\bf Howard Baer}, University of Oklahoma;
    {\bf Vernon Barger}, University of Wisconsin;
    {\bf Laura Baudis}, University of Z\"urich;
    {\bf Daniel Bauer}, Fermilab;
    {\bf James Buckley}, Washington University; 
    {\bf Matthew Cahill-Rowley}, SLAC; 
    {\bf Randel Cotta}, University of California, Irvine; 
    {\bf Alex Drlica-Wagner}, SLAC;
    {\bf Jonathan L.~Feng}, University of California, Irvine;
    {\bf Stefan Funk}, SLAC;
    {\bf JoAnne Hewett}, SLAC; 
    {\bf Dan Hooper}{\footnote{Corresponding authors: {\tt dhooper@fnal.gov}, {\tt  mkapling@uci.edu}, 
    {\tt matchev@phys.ufl.edu}, and {\tt ttait@uci.edu}}}, Fermilab;
    {\bf Ahmed Ismail}, SLAC; 
    {\bf Manoj Kaplinghat}{${}^\ast$},  University of California, Irvine;
	{\bf Kyoungchul Kong}, University of Kansas;
    {\bf Alexander Kusenko},  University of California, Los Angeles;
    {\bf Konstantin Matchev}{${}^\ast$}, University of Florida; 
	{\bf Mathew McCaskey}, University of Kansas;
    {\bf Daniel McKinsey}, Yale University;
    {\bf Dan Mickelson}, University of Oklahoma;
    {\bf Tom Rizzo}, SLAC; 
	{\bf David Sanford}, Caltech;
	{\bf Gabe Shaughnessy}, University of Wisconsin;
    {\bf William Shepherd}, University of California, Santa Cruz;
    {\bf Tim M.~P.~Tait}{${}^\ast$}, University of California, Irvine;
    {\bf Xerxes Tata}, University of Hawaii;
    {\bf Sean Tulin}, University of Michigan;
    {\bf Alexander M. Wijangco}, University of California, Irvine; 
    {\bf Matthew Wood}, SLAC;
    	{\bf Jonghee Yoo}, Fermilab;
    {\bf Hai-Bo Yu}, University of California, Riverside;
    \\
{\it on behalf of the Snowmass 2013 Cosmic Frontier WG4 "Dark Matter Complementarity"}\\
Conveners: {\bf Dan Hooper}, {\bf Manoj Kaplinghat}, {\bf Konstantin Matchev}\\ 
}

\vspace{1cm}
\date{30 October 2013}

\noaffiliation
%\email{name@place.edu}
%\affiliation{Place
%\PRE{\vspace*{.4in}}
%}

\begin{abstract}
In this Report we discuss the four complementary searches for the identity of dark matter: direct
detection experiments that look for dark matter interacting in the
lab, indirect detection experiments that connect lab signals to dark
matter in our own and other galaxies, collider experiments that elucidate
the particle properties of dark matter, and astrophysical probes
sensitive to non-gravitational interactions of dark matter. 
The complementarity among the different dark matter searches is discussed qualitatively 
and illustrated quantitatively in several theoretical scenarios. 
Our primary conclusion is that the diversity of possible 
dark matter candidates requires a balanced program based on all 
four of those approaches.

%In this Report we summarize the many dark matter searches currently
%being pursued in four complementary approaches: direct
%detection experiments that look for dark matter interacting in the
%lab, indirect detection experiments that connect lab signals to dark
%matter in our own and other galaxies, collider experiments that elucidate
%the particle properties of dark matter, and astrophysical probes
%sensitive to non-gravitational interactions of dark matter. 
%The essential features of broad classes of experiments are described,
%each with their own strengths and weaknesses. The complementarity 
%among the different dark matter searches is discussed qualitatively 
%and illustrated quantitatively in several theoretical scenarios. 
%Our primary conclusion is that the diversity of possible 
%dark matter candidates requires a balanced program based on all 
%four of those approaches.
\end{abstract}

%\pacs{95.35.+d}
%%95.35.+d Dark matter

\maketitle

\footnotesize
\tableofcontents
\normalsize

\section{Introduction}
\label{sec:intro}

Despite being five times as abundant as normal matter in the 
Universe, the identify of dark matter is unknown. 
Its existence, however, implies that our inventory of the basic building blocks of nature is
incomplete, and uncertainty about its properties clouds attempts to
fully understand how the Universe evolved to its present state and how
it will evolve in the future.  Uncovering the identity of dark matter is therefore a central and grand
challenge for both fundamental physics and astronomy. Fortunately, 
a very promising array of groundbreaking experiments are positioned to transform the field of
dark matter in the coming decade.  The prospect that dark matter particles might be observed in the near future has drawn many new researchers to the field, which is now characterized by an
extraordinary diversity of approaches unified by the common goal of
discovering the identity of dark matter.

Dark matter was first postulated in its modern form in the 1930s to
explain the anomalously large velocities of galaxies in the Coma
cluster~\cite{Zwicky:1933gu}.  Over the subsequent decades, evidence for dark matter grew to include data from galactic rotation curves~\cite{Rubin:1970zz,Rubin:1980zd,1978PhDT.......195B}, and more recently from weak~\cite{Refregier:2003ct} and strong~\cite{Tyson:1998vp} lensing,
hot gas in clusters~\cite{Lewis:2002mfa,Allen:2002eu}, the Bullet
Cluster~\cite{Clowe:2006eq}, Big Bang nucleosynthesis
(BBN)~\cite{Fields:2008}, 
%further constraints from large scale structure~\cite{Allen:2002eu}, 
distant supernovae~\cite{Riess:1998cb,Perlmutter:1998np}, 
the statistical distribution of galaxies~\cite{Tegmark:2003ud,Hawkins:2002sg}
and the cosmic microwave background (CMB)~\cite{Komatsu:2010fb,Ade:2013lta}.
Together, these data provide an overwhelming body of evidence in support of the conclusion that cold (or possibly warm) dark matter makes up roughly a quarter of the total energy density of the Universe, exceeding the density of normal matter by about a factor of five. 

%All of this evidence for dark matter derives from its gravitational
%pull on visible matter.  This does little to shed light on the
%identity of dark matter, since all particles interact universally
%through gravity.  To make progress, dark matter must be detected
%through non-gravitational interactions. There are many possibilities.
%For reviews, see, \eg,
%Refs.~\cite{Bertone,Bergstrom:2009ib,Feng:2010gw}.

Although these observations strongly support the existence of dark matter, they each do so uniquely through the dark matter's gravitational influence on visible matter. As a consequence, these observations do little to shed light on the particle identity of dark matter, since all forms of matter interact universally
through the force of gravity.  If dark matter's nature is to be understood, it is necessary that it  be detected
through non-gravitational interactions. Possibilities for such observations include the elastic scattering of dark matter with nuclei in a detector (direct detection), the detection of standard model particles produced in the annihilations or decays of dark matter (indirect detection), the production of dark matter particles at colliders, and the effects of dark matter interactions on astrophysical systems. 

The prospects for each of these observational approaches, of course, depend considerably on the dark matter candidate under consideration, for which there are very many possibilities (for reviews, see, \eg, Refs.~\cite{Bertone,Bergstrom:2009ib,Feng:2010gw}). In the case of weakly interacting massive particles (WIMPs), dark
matter particles are expected to have been produced in the hot early Universe and then
annihilate in pairs.  Those that survive to the present are known as
``thermal relics''~\cite{Zeldovich:1965,Chiu:1966kg,Steigman:1979kw,Scherrer:1985zt}. The realization that a stable particle species with a GeV-TeV mass and an annihilation cross section near the scale of the weak interaction will yield a thermal relic abundance similar to the observed cosmological abundance of dark matter ($\Omega_X \sim {\cal O}(0.1)$) has provided a great deal of motivation for dark matter in the form of WIMPs. Furthermore, WIMPs are generically predicted in a wide variety of weak-scale extensions of
the standard model, including models with
supersymmetry~\cite{Goldberg:1983nd,Ellis:1983ew}, extra spatial
dimensions~\cite{Servant:2002aq,Cheng:2002ej,Agashe:2004bm,Cembranos:2003mr}, and many others~\cite{BirkedalHansen:2003mpa,Cheng:2003ju,LopezHonorez:2006gr}.

In addition to WIMPs, there are many other well motivated and often studied candidates for dark matter.
Axions~\cite{Peccei:1977ur,Weinberg:1977ma,Wilczek:1977pj,Asztalos:2006kz}
are motivated by the fact that, despite observations to the contrary, the strong interaction of the standard model is naturally expected to induce significant CP violating effects. Although the Pecci-Quinn mechanism is capable of elegantly solving this problem, it also predicts the existence of a very light and extremely weakly interacting axion, which could make up the observed dark matter abundance.  Right-handed or sterile neutrinos are motivated by the observation of 
non-zero neutrino masses, and for certain ranges of masses and interaction strengths,
may also be viable candidates for dark matter~\cite{Dodelson:1993je,Kusenko:2009up,Abazajian:2012ys}.
Alternatively, dark matter may be in a so-called hidden sector, which
has its own set of matter particles and forces through which the dark
matter interacts with other currently unknown particles \cite{Feng:2009mn,Kaplan:2009de}.  Another possibility is that the dark matter may be asymmetric~\cite{Nussinov:1985xr,Gelmini:1986zz,Kaplan:1991ah,Barr:1991qn,Kaplan:2009ag}, with a slight imbalance between the numbers of dark particles and dark antiparticles in the early universe.  These particles annihilate until only the
slight excess of dark particles remains.  In many models, the dark
matter asymmetry is related to the baryon-antibaryon asymmetry, and one expects the number of dark matter particles in the Universe to be similar to the number of baryons.  Since dark matter contributes roughly 
five times more to the energy density of the Universe than normal matter, 
this scenario predicts dark matter particles with a mass on the order of $\sim 1 - 15~\gev$.

%Although these dark matter candidates differ in important ways, in
%most cases, they have non-gravitational interactions through which they
%may be detected.
%%\footnote{A notable exception is primordial black holes~\cite{Carr:2003bj,Frampton:2010sw}.}  
%The non-gravitational interactions may be with any
%of the known particles or, as noted above for hidden sector dark
%matter, with other currently unknown particles.  These possibilities
%are shown in \figref{interactions}, where the particles are grouped
%into four categories: nuclear matter; leptons; photons and other
%bosons; and other as-yet unknown particles.  Dark matter may interact
%with one type of particle, or it may interact with several. 

\begin{figure}[tb]
\includegraphics[width=0.95\columnwidth]{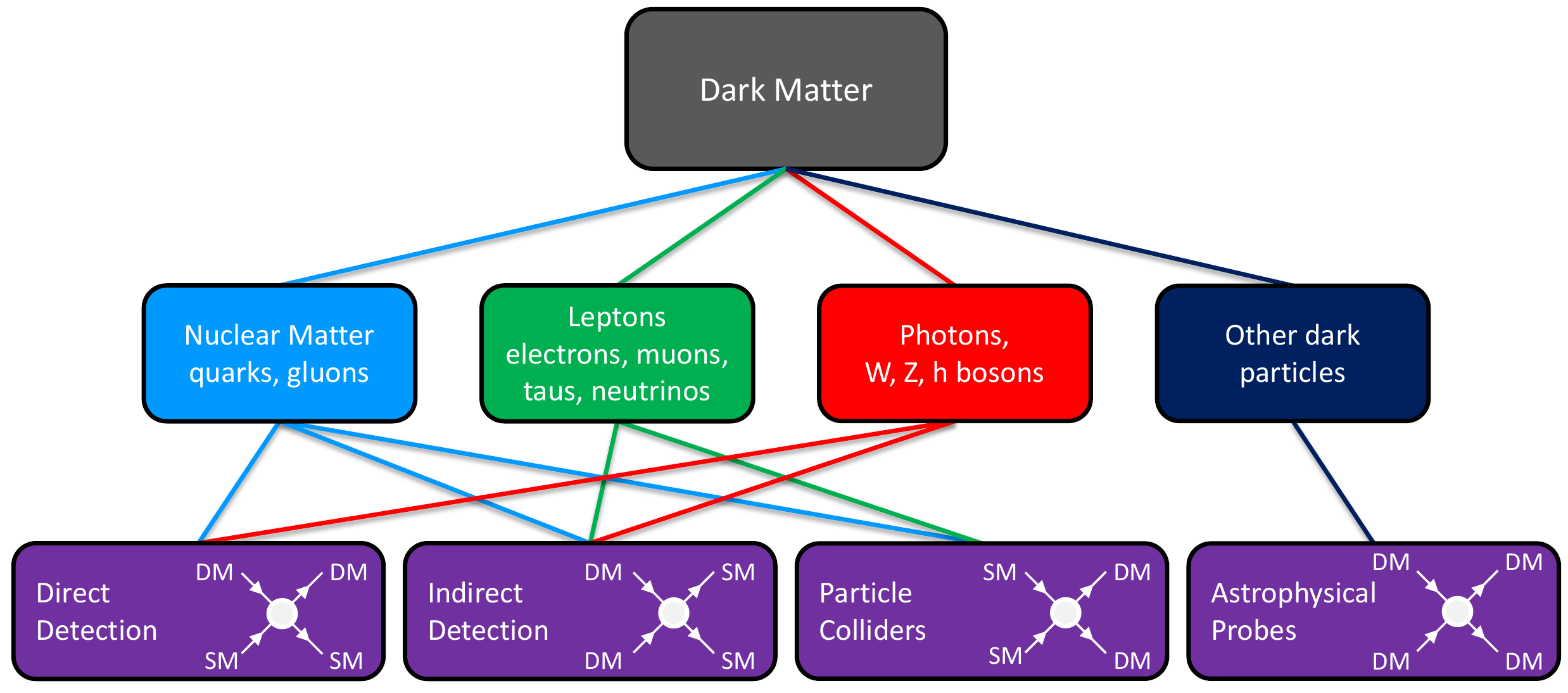}
%\vspace*{-.1in}
\caption{Dark matter may have non-gravitational interactions with one
  or more of four categories of particles: nuclear matter, leptons,
  photons and other bosons, and other dark particles.  These
  interactions may then be probed by four complementary approaches:
  direct detection, indirect detection, particle colliders, and
  astrophysical probes. The lines connect the experimental approaches
  with the categories of particles that they most stringently probe 
  (additional lines can be drawn in specific model scenarios).
  The diagrams give example reactions of dark matter with
  standard model particles (SM) for each experimental approach.
\label{fig:interactions}}
\end{figure}

Although these dark matter candidates differ in important ways, in
most cases they have non-gravitational interactions through which they
may be detected. The non-gravitational interactions may be with any
of the known particles or, as noted above for hidden sector dark
matter, with other currently unknown particles.  These possibilities
are illustrated in \figref{interactions}, where the particles are grouped
into four categories: nuclear matter, leptons, photons and other
bosons, and other as-yet unknown particles.  Dark matter may interact
with one type of particle, or it may interact with several. 

As we will discuss, discovering the particle identify of dark matter will most likely require synergistic progress along many different lines of inquiry.  Our
primary conclusion is that the diversity of possible dark matter
candidates, and the diversity of their observational and experimental signatures, motivates a broad and well-balanced program of research. In particular, such a program requires a diverse array of experiments, together covering each of what we will refer to as the four pillars of dark matter discovery: 
\begin{itemize}
\setlength{\itemsep}{1pt}\setlength{\parskip}{0pt}\setlength{\parsep}{0pt}
\item {\em Direct Detection}.  Dark matter scatters off a detector,
  producing a detectable signal.  Prime examples are the detection of
  WIMPs through scattering off nuclei and the detection of axions
  through their interaction with photons in a magnetic field.
\item {\em Indirect Detection}.  Pairs of dark matter particles
  annihilate producing high-energy particles (antimatter, neutrinos,
  or photons).  Alternatively, dark matter may be metastable, and its
  decay may produce the same high-energy particles.
\item {\em Particle Colliders}.  Particle colliders, such as the Large
  Hadron Collider (LHC) and proposed future colliders, produce
  dark matter particles, which escape the detector, but are
  discovered as an excess of events with missing energy or momentum.
\item {\em Astrophysical Probes}.  The particle properties of dark
  matter are constrained through its impact on astrophysical
  observables.  In particular, dark matter's non-gravitational interactions could observably impact the densities of dark matter present in the central regions of galaxies, or the amount of dark matter substructure found in halos. Such interactions may also alter the cooling rates of stars, and influence the pattern of temperature fluctuations observed in the cosmic microwave background.
\end{itemize}

These search strategies are each shown in \figref{interactions} and are
connected to the particle interactions that they {\em most stringently} probe.

After summarizing many of the most promising particle candidates for dark matter in Sec.~\ref{sec:candidates}, we return in Sec.~\ref{sec:probes} to these four pillars in more detail, discussing the current status and future prospects of direct, indirect, and collider searches for dark matter, as well as the impact of astrophysical observations. In Sec.~\ref{sec:basic} we begin the discussion of the complementarity between different dark matter search strategies at a qualitative level. We extend this further in Sec.~\ref{sec:quantitative}, discussing quantitatively the interplay between experimental approaches, considering a number of representative particle physics frameworks.  Finally, we summarize our conclusions in Sec.~\ref{sec:conclusions}.

\section{Dark Matter Candidates}
\label{sec:candidates}

In this section, we briefly summarize a number of specific dark matter candidates and candidate classes that have been considered in the literature. While certainly not exhaustive, this discussion is intended to reflect a representative sample of how the particle physics community currently views the form that dark matter particles might take.

%\subsection{WIMP candidates}

\subsubsection{Neutralinos and other Supersymmetric WIMPs}

Models with weak-scale supersymmetry are motivated by a number of theoretically attractive features, including their ability to solve the electroweak hierarchy problem~\cite{hierarchy1,Veltman:1980mj,Witten:1981nf,Kaul:1981wp}, allow for the high-scale unification of the standard model gauge couplings~\cite{Dimopoulos:1981zb,Dimopoulos:1981yj,Sakai:1981gr,Ibanez:1981yh,Einhorn:1981sx}, and provide a viable candidate for the dark matter of our Universe~\cite{Goldberg:1983nd,Ellis:1983ew}. In particular, in models with unbroken $R$-parity, the lightest supersymmetric particle (LSP) is stable. If the LSP is a neutralino (a mixture of the superpartners of the neutral gauge and Higgs bosons), it will be a WIMP. Sneutrinos could also consitute a WIMP, although constraints from direct detection experiments exclude this possibility in minimal models. Gravitinos could also be the LSP, but in this case the dark matter would far less than weakly interacting and would require a non-thermal origin in the early Universe (see below).

In recent years, null results from the LHC and direct detection experiments have begun to significantly constrain the parameter space of most commonly studied supersymmetric models. However, neutralino dark matter remains viable as a thermal relic in a number of regions of parameter space, including those in which its depletion in the early Universe occurs through coannihilations with another species (such as a stau or stop), or through the neutral heavy Higgs resonance. It is also possible that the LSP is a mixture of gauginos and higgsinos ({\it e.g.} focus point models), or is a relatively heavy wino ($m_{\chi}\approx2.7$-3 TeV) or higgsino ($m_{\chi}\approx1$ TeV).

\subsubsection{Dark Matter from Extra Dimensions }

If there exist dimensions of space beyond the three we directly experience, Kaluza-Klein (KK) excitations of standard model particles could potentially make up the dark matter of the universe. In models with one (or more) universal, flat dimension, KK parity (a remnant of n-dimensional momentum conservation) can stabilize the lightest KK state. Of particular interest is the lightest KK excitation of the hypergcharge gauge boson~\cite{Servant:2002aq,Cheng:2002ej}, which for a $\sim$TeV scale mass leads to a thermal relic abundance similar to the observed dark matter abundance. KK dark matter candidates in other scenarios have also been proposed, including those arising in scenarios with warped extra dimensions~\cite{Agashe:2004ci,Agashe:2004bm}.

\subsubsection{Axions}

In the standard model, the naive expectation for the neutron's electric dipole moment is roughly $10^{12}$ times larger than is allowed experimentally. It was recognized by Peccei and Quinn, however, that this ``strong CP problem'' could be solved by introducing a spontaneously broken global $U(1)$ symmetry which dynamically suppresses the CP violation induced by QCD~\cite{Peccei:1977ur}. The Goldstone boson of this broken global symmetry is the axion, which acquires a mass from the QCD anomaly~\cite{Wilczek:1977pj,Weinberg:1977ma}. In light of experimental and astrophysical constraints, the scale of this symmetry breaking must be very high, implying that the axion will be very light ($m_{a} \lsim 0.01$~eV) and extremely feebly interacting. Such particles never reach thermal equilibrium in the early Universe, but are predicted to be produced as cold dark matter through non-thermal mechanisms (such as through vacuum misalignment).

Efforts to probe the parameter space in which axions could constitute the dark matter of our Universe rely on the photon-photon-axion coupling~\cite{Sikivie:1983ip}. In particular, microwave cavity experiments utilizing this interaction have begun to constrain a small fraction of such models, and are well-positioned to test the a much larger fraction of the axion dark matter window in the relatively near future.

\subsubsection{Gravitinos}

If the lightest supersymmetric particle is the superpartner of the graviton, it could also constitute the dark matter. Unlike the case of neutralinos or other WIMPs, the extremely feeble interactions of the gravitino prevent it from reaching thermal equilibrium with the thermal bath present in the early Universe. As a consequence, the surviving density of gravitinos depends strongly on the temperature to which the Universe was reheated following inflation. 

The mass of the gravitino can be constrained by requiring that the decays of heavier superpartners do not destroy the successful predictions of big bang nucleosynthesis or overly alter the characteristics of the cosmic microwave background. If the gravitino is very light ($\lsim 100$ MeV), however, the heavier superpartners will be sufficiently short lived as to decay prior to big bang nucleosynthesis, thus evading these constraints~\cite{Pagels:1981ke,PhysRevLett.48.1303}. Alternatively, if the gravitino mass is of the same order of magnitude as that of the next-to-lightest superpartner (presumably near the electroweak-scale), such decays can occur after the formation of light elements, but early enough to not unacceptably distort the cosmic microwave background. In this later case, the gravitino dark matter could originate from the decays of heavier superpartners, in what is known as the ``superWIMP'' scenario~\cite{Feng:2003xh,Feng:2003uy}.

%\subsection{Non-WIMP candidates}

\subsubsection{Asymmetric Dark Matter}

In asymmetric dark matter (ADM) scenarios, the dark matter relic abundance today is linked to a primordial dark asymmetry generated in the early Universe.  The situation is analogous to how the baryon density arises: the Universe carries a net baryon number, such that when baryons and antibaryons annihilate around the QCD epoch, the residual asymmetric density of baryons is left over.  For ADM, an initial asymmetry between dark matter particles $\chi$ and antiparticles $\bar \chi$ (assuming $\chi \ne \bar \chi$) leads to a relic density of $\chi$ only, while the symmetric $\chi$-$\bar \chi$ density annihilates away.  Examples of ADM candidates include technibaryons~\cite{Nussinov:1985xr,Barr:1990ca}, scalar neutrinos~\cite{Hooper:2004dc}, or exotic fermions~\cite{Kaplan:1991ah}, as well as a wide variety of other possibilities (see Refs.~\cite{Davoudiasl:2012uw,Petraki:2013wwa} for reviews).  The baryon and dark matter asymmetries may be related, e.g., through higher-dimensional operators~\cite{Kaplan:2009ag}, offering a unified origin for the generation of both dark and visible matter.  The cosmic ratio of dark matter-to-baryons is given by $\Omega_{\rm DM}/\Omega_b \sim m_\chi/m_p$, where $m_p$ is the proton mass.  ADM can explain the apparent coincidence between the similar energy densities of dark and visible matter in the Universe ($\Omega_{\rm DM}/\Omega_b \sim 5$) if the dark matter mass is in the range $m_\chi \sim 1-20$ GeV.  Another requirement is that the $\chi \bar \chi$ annihilation cross section must be larger than for a typical WIMP to annihilate away the symmetric density.  Indirect detection signals through annihilation today are typically absent, since only $\chi$ is present in dark matter halos today. However, accumulation of ADM in old stars can lead to observable effects~\cite{McDermott:2011jp,Kouvaris:2011fi,Zentner:2011wx,Iocco:2012wk}. ADM can also produce other exotic signatures; e.g., ADM in the local halo may annihilate with visible baryons, producing signals in nucleon decay searches~\cite{Davoudiasl:2010am}.

\subsubsection{Non-thermal WIMPs}

It is possible that the dark matter might not be a thermal relic of the early universe, but was instead produced through a non-thermal mechanism. For example, decays of relatively long-lived stable species could produce WIMPs well after they would have ordinarily frozen out of equilibrium. Although neutralinos produced via the decay of moduli~\cite{Moroi:1999zb} or Q-balls~\cite{Fujii:2002kr} represent well-studied examples of non-thermal WIMP production, many other scenarios are also possible (e.g.,~\cite{Allahverdi:2012gk,Allahverdi:2013noa}).

\subsubsection{Hidden Sector Dark Matter}
In hidden sector dark matter scenarios, the dark matter does not interact through the forces of the standard model, but is instead charged under new ``dark'' gauge symmetries. In the early Universe, dark gauge bosons provide a thermal bath and a hidden sector experiences its own thermal history, distinct from that of the visible sector~\cite{Feng:2008mu}. 
%Within the context of supersymmetry, hidden dark matter can still keep virtues of WIMPs, namely, the connection to the gauge hierarchy problem~\cite{Feng:2008ya}. 
Hidden sector dark matter can have a rich structure with interesting features that usual WIMPs do not carry. In many hidden sector dark matter models, dark gauge forces mediate dark matter strong self-interactions that are relevant for astrophysics~\cite{Feng:2009mn,Ackerman:2008gi,Feng:2009hw,Buckley:2009in,Loeb:2010gj}. Dark radiation can delay dark matter kinetic decoupling and suppress the matter power spectrum in small scales~\cite{Feng:2009mn, CyrRacine:2012fz}. In contrast to the WIMP paradigm, hidden sector dark matter candidates are not limited to fundamental particles, but could instead consist of composite states, such as hidden atoms~\cite{Kaplan:2009de,CyrRacine:2012fz} or mirror baryons~\cite{Foot:2004pa,An:2009vq}. The dark matter could also be made up of multiple components, with a small fraction being dissipative and forming a dark disk~\cite{Fan:2013yva}. These novel features can leave observable signatures on structure formation by modifying the internal structure of halos, as well as the number of halos of a given mass (for the case with dark radiation). In addition, hidden sector dark matter could also couple to the standard model in the presence of a connector sector, such as the Higgs portal~\cite{Patt:2006fw}, kinetic mixing~\cite{Holdom:1985ag}, or mass mixing of gauge bosons~\cite{Feldman:2006wd,Frandsen:2011cg}. In such cases, hidden sector dark matter provide signals in direct and indirect detection experiments~\cite{ArkaniHamed:2008qn,Pospelov:2008jd,Feldman:2008xs}.

\section{The Four Pillars of Dark Matter Detection}
\label{sec:probes}

\subsection{Direct Detection}
\label{sec:direct}

%{\bf Include figures from CF1 report summarizing the current limits.}

\subsubsection{Scattering of WIMPs on nuclei}

Dark matter permeates the whole Universe, and its local density on
Earth is known to be $7\times 10^{-25}~\g/\cm^3$ (0.4 GeV/cm$^3$) to within a factor of 2.
This creates the opportunity to detect dark matter particles {\em
  directly} as they pass through and scatter off normal
matter~\cite{Goodman:1984dc}.  Such events are extremely rare, and so
the direct detection approach requires sensitive detectors with
exquisite background rejection.  The expected signals and prospects for discovery depend on the
nature of the dark matter particles and their interactions.  For a
list of current and planned experiments, see Ref.~\cite{TableDD}.

If the dark matter consists of WIMPs, direct searches are extremely promising. It has long been appreciated that if the WIMPs' coupling to quarks is chosen such that the resulting thermal relic abundance is consistent with the observed dark matter density, the same couplings would lead to potentially detectable scattering cross sections with nuclei~\cite{Goodman:1984dc}. As the sensitivity of direct detection experiments has improved, many otherwise viable dark matter candidates have been ruled out. Of particular interest are WIMPs in the form of a scalar or a Dirac fermion which annihilates primarily though a coupling to the $Z$ (such as a sneutrino, KK neutrino, or stable fourth generation neutrino). Such dark matter candidates are currently ruled out by direct detection constraints by multiple orders of magnitude. Excitingly, WIMPs which annihilate largely through Higgs exchange in the early universe (including neutralinos in many supersymmetric models) predict direct detection rates which are near the sensitivity of current experiments.

Experimental techniques employed in direct searches include detectors that record combinations of ionization,
scintillation light, and phonons (heat).  The most sensitive detectors
employ multiple techniques, enabling for discrimination between dark matter scattering events and backgrounds. Depending on the target material,
direct detection experiments are sensitive to a combination of spin-dependent and
spin-independent interactions between dark matter and nuclei. The sensitivity of the
current generation of detectors for spin-independent scattering with protons or neutrons is approaching $\sigma_{\text{SI}}^{p,n} \sim
10^{-45}~\cm^2$ for WIMP masses of $\sim$$100~\gev$. The sensitivity of the leading direct detection experiments is expected to further improve by multiple orders of magnitude over the coming decade.  Although direct detection is more difficult for dark matter particles with $\sim$1-10 GeV, there has recently been significant progress in developing experiments with low threshold energies. More details on the current and planned experiments can be found in the Cosmic Frontier subgroup 1 (CF1) report.

%%%

%%%

\subsubsection{Direct Axion Searches}

Axions also have strong prospects for direct detection. Cosmological
and astrophysical constraints restrict the allowed axion mass range to
be between approximately 1 $\mu$eV and 1 meV.  In a static magnetic field, there is
a small probability for cosmologically produced axions to be converted
by virtual photons to real microwave photons by the Primakoff
effect~\cite{Sikivie:1983ip}. This would produce a monochromatic
signal with a line width of $dE/E\sim 10^{-6}$, which could be detected
in a high-$Q$ microwave cavity tunable over GHz frequencies.  
In the near future, such searches will be sensitive to models with 
axion mass $\sim \mu$eV, which is the favored region if axions 
are to constitute a significant component of the dark matter. More details can be found in the reports of the Cosmic Frontier subgroups 1 and 3. 

\subsection{Indirect Detection}
\label{sec:indirect}

%\subsubsection{Energetic Particles From Dark Matter Annihilations or Decays}

Even if WIMPs are stable in isolation, they may undergo pair annihilation, producing energetic photons, neutrinos, and cosmic rays. Alternatively, if the dark matter is unstable, their decays could produce energetic standard model particles. A wide variety of efforts to detect such annihilation or decay products are currently underway (current and planned indirect search experiments are listed in~\cite{TableID}). We provide a short summary of the methods below to set the stage for the discussion on complementarity. More details can be found in the Cosmic Frontier subgroup 2 (CF2) report.

The most stringent and broadly applicable constraints on the dark matter annihilation cross section (or lifetime) have been derived from gamma-rays observations from the Fermi Gamma-Ray Space Telescope (FGST) and ground based gamma-ray telescopes.  To date, Fermi data has been used to search for dark matter annihilation products from dwarf spheroidal galaxies~\cite{GeringerSameth:2011iw,Ackermann:2011wa} and the Galactic Center~\cite{Hooper:2012sr} (as well less restrictive constraints from observations of galaxy clusters, the Galactic Halo, galactic subhalos, and the isotropic gamma-ray background). These constraints are beginning to probe annihilation cross sections at or around the value predicted for a simple thermal relic ($\sigmath v \sim 3 \times 10^{-26}$ cm$^3$/s), at least for dark matter particles with masses below a few tens of GeV and which annihilate to final states which result in significant fluxes of gamma-rays. And while the particle (or particles) that make up the dark matter certainly could possess an annihilation cross section that is well below this value, the range of cross sections that is presently being probed by Fermi represents a very well motivated and theoretically significant benchmark. Although the strongest current constraints for significantly heavier dark matter candidates, (provided by ground based telescopes~\cite{Abramowski:2013ax,Abramowski:2010aa,VERITAS:2013fra,Aliu:2012ga,Aleksic:2011jx}) are not yet at the sensitivity to probe this benchmark value, the future CTA array is expected to be able to reach this level of sensitivity for WIMPs with masses between $\sim 200$ GeV and several TeV. 

Searches for dark matter annihilation or decay products in the cosmic ray spectrum has received a great deal of attention in recent years, motivated in large part by the unexpectedly high flux of cosmic ray positrons observed by the PAMELA experiment~\cite{Adriani:2008zr} (and now confirmed with high precision by AMS~\cite{Aguilar:2013qda}). And although dark matter models have been proposed that could account for this signal, the observed positrons may also originate from nearby pulsars or other astrophysical objects~\cite{Hooper:2008kg}. Even if dark matter is not responsible for the large high-energy cosmic ray positron flux, however, the lack of sharp features in this spectrum can be used to place fairly stringent constraints on the dark matter annihilation cross section to electron-positron or muon-antimuon pairs~\cite{Bergstrom:2013jra}.  In this respect, cosmic ray positron measurements are complementary to gamma-ray searches for dark matter, which are most sensitive to other annihilation channels.  Measurements of the cosmic ray antiproton spectrum~\cite{Adriani:2010rc} are also sensitive to the annihilating dark matter at a level that is competitive with existing gamma-ray constraints. When cosmic ray and gamma-ray measurements are combined with multi-wavelength observations, such as those from radio and X-ray telescopes, it can provide a powerful means of disentangling would-be dark matter signals from astrophysical backgrounds. 

Although dark matter annihilations in the galactic halo produce too few neutrinos to be detected in most models, annihilations which occur in the center of the Sun could potentially generate an observable flux of high energy neutrinos. Dark matter particles scatter with nuclei and become captured in the Sun at a rate determined by the WIMP's elastic scattering cross section. In many models, equilibrium can be reached between the capture and annihilation rates; in this case, the resulting neutrino flux does not depend on the annihilation cross section of the dark matter. For this reason, neutrino telescopes probe very different characteristics of the dark matter than other indirect search strategies. In the case in which the dark matter scatters with nuclei largely through spin-dependent couplings, the current constraints from the IceCube experiment are competitive with direct detection searches~\cite{Aartsen:2012kia}. The Sun is also a potential source of dark matter axions produced in the Primakoff conversion of plasma photons.
Such solar axions could be detected when they re-convert in the magnetic field of a detector on the Earth \cite{Arik:2011rx}.

Dark matter annihilations taking place during the epoch of reionization could potentially produce a sufficient number of energetic particles to observably impact the anisotropies of the cosmic microwave background~\cite{Padmanabhan:2005es,Mapelli:2006ej,Zhang:2006fr}. Measurements from WMAP and Planck provide constraints that are only modestly weaker than those derived from gamma-ray observations, and without many of the astrophysical uncertainties associated with such approaches~\cite{Slatyer:2009yq}.

The various indirect detection strategies described in this section are in many ways highly complementarity, potentially providing different information, being sensitive to different dark matter candidates, and suffering from different astrophysical uncertainties. For example, while interpretations of cosmic ray positron measurements rely on models of the Milky Way's magnetic field, radiation field, and gas distributions, gamma-ray observations are not sensitive to such factors (gamma-rays travel without deflection, and with negligible energy losses over Galactic scales). Gamma-ray constraints on dark matter annihilation do, however, depend on the dark matter distributions in the inner regions of halos, and on our ability to understand or model the relevant astrophysical backgrounds. Similarly, while positron measurements are very sensitive to dark matter particles which annihilate directly to electron-positron pairs but are far less sensitive to scenarios in which the WIMPs annihilate to quarks, the inverse is true for gamma-ray searches.

\subsection{Particle Colliders}
\label{sec:colliders}

%{\bf Include figures from CMS and ATLAS.}

Dark matter may also be produced in high-energy particle collisions.
For example, if dark matter has substantial couplings to nuclear
matter, it can be created in proton-proton collisions at the
Large Hadron Collider (LHC).  Once produced, dark matter particles
will likely pass through detectors without a trace, but their
existence may be inferred from an imbalance in the visible momentum,
just as in the case of neutrinos.  Searches for dark matter at the LHC
are therefore typified by missing momentum, and can be categorized by
the nature of the visible particles that accompany the dark matter
production.  Because backgrounds are typically smaller for larger
values of missing momentum, collider searches tend to be most
effective for low-mass dark matter particles, which are more easily
produced with high momentum.

There are two primary mechanisms by which the LHC could hope to
produce dark matter together with hadronic jets (see, e.g., Chapters
13 and 14 of Ref.~\cite{Bertone}).  In the first, two
strongly interacting parent particles of the dark matter theory are
produced, and each one subsequently decays into the dark matter and
standard model particles, resulting in missing momentum plus two or
more jets of hadrons.  Since the production relies on the strong
force, the rate of production is specified by the color charge, mass,
and spin of the parent particles and is typically rather insensitive
to the mass of the dark matter itself.  Current null results from LHC
searches for the supersymmetric partners of quarks exclude such
particles with masses less than $\sim 1.5~\tev$.

A second mechanism produces the dark matter directly
together with additional radiation from the initial quarks or gluons
participating in the reaction, resulting in missing momentum recoiling
against a single ``mono-jet.''
Since this process does not rely as
explicitly on the existence of additional colored particles that
decay into dark matter, it is somewhat less sensitive to the details
of the specific theory and places bounds directly in the parameter
space of the dark matter mass and interaction strength.  However, one
does need to posit a specific form of the interaction between the dark
matter and quarks or gluons.  For electroweak-size couplings and
specific choices of the interaction structure, these searches exclude
dark matter masses below about 500 GeV.

High energy lepton colliders may create dark matter through analogous
processes, such as production of dark matter along with a photon
radiated from the initial leptons.  For electroweak-size couplings of
dark matter to electrons, LEP excluded dark matter masses below about
90 GeV.  A future high-energy lepton collider could conceivably
discover dark matter particles with masses up to roughly half the
collision energy, \eg, 500 GeV for a 1 TeV ILC.  For a list of current
and proposed future colliders, see~\cite{TableEF}.

\subsection{Astrophysical Probes}
\label{sec:astrophysics}

Many dark matter candidates, such as neutralinos, are cold and effectively collisionless (i.e., gravitational interactions dictate the resulting large scale structure). The adjective ``cold" implies that the temperature of dark matter at formation is such that it allows for structure to form with masses orders of magnitude below that of the smallest galaxies observed. Predictions for cold, collisionless dark matter (CDM) agree very well with cosmological data~\cite{Komatsu:2010fb}, but CDM may be an approximation that breaks down on small (galactic and sub-galactic) sales. Thus, observations on galactic scales provide an opportunity to detect or constrain the possibility of dark matter that is warm or strongly self-interacting. 

In recent years, observations have suggested that the central densities of some dark matter halos may be lower than expected. In particular,  evidence for low central densities has been seen in a variety of self-gravitating systems, including satellite galaxies
of the Milky Way, spiral galaxies, and clusters of galaxies. (See section~\ref{sec:self} for a short discussion.) Although such low density cores naively appear to be in conflict with the simplest predictions for cold, collisionless dark matter, it has been argued that baryonic feedback (such as outflows from supernovae, or the environment of galaxies) may be able to reconcile such disparities (see {\em e.g.}, \cite{Weinberg:2013aya}). Alternatively, these observations may be indicating that the dark matter is not entirely cold and collisionless, but may instead be either warm or strongly self-interacting.

A key prediction of collisional and collisionless cold dark matter is the presence of thousands or more dark subhalos (self-bound clumps of dark matter) within the halo of galaxies like the Milky Way \cite{Klypin:1999uc,Moore:1999nt,Rocha:2012jg}.  In order to confirm a candidate like the WIMP as being all of dark matter, it is essential to verify this prediction. A clear lack of the expected number of dark subhalos would show that the dark matter is ``warm." Compared to cold dark matter, models of warm dark matter predict less power in small-scale density fluctuations, dramatically reducing the predicted number of low-mass dark matter halos.
Hidden sector models in which the dark matter interacts with other light particles can also lead to similar effects. The mass-scale below which halo formation is suppressed is directly related to one or more parameters of the particle physics model --- for example, mass and couplings of sterile neutrino dark matter 
candidates~\cite{Kusenko:2006rh}.  In addition, although the central densities of dark matter halos are reduced in warm dark matter 
cosmology, large constant density cores have not been shown to form. 

The key prediction of thousands of cold dark matter subhalos can be tested in the future using strong gravitational lensing systems, precise observations of the clustering in the universe and searches for new satellite galaxies and stellar streams.  Proposed future observatories such as LSST~\cite{Abell:2009aa} and WFIRST~\cite{Spergel:2013tha} will be able to make decisive advances towards testing this prediction, with the following expected goals. 
\begin{itemize}
\item A deep wide-field survey should lead to the discovery of new ultra-faint satellites of the Milky Way. 
\item A high latitude survey will allow proper motions of halo stars at distances of 100 kpc to measured, which will allow the mass of the Milky Way to be estimated better. This is important because issues with small-scale structure based on arguments about Milky Way's satellite population depend sensitively on the mass of the Milky Way. 
\item Measuring proper motions of stars in streams associated with globular clusters like Pal 5 created by the tides of the Milky Way will allow the gaps in these streams to be characterized. Dark subhalos passing through streams will create gaps and the nature of these gaps could reveal the presence of subhalos down to a million solar masses, for which there are no visible counterparts~\cite{Carlberg:2012ur}.
\item Huge increase in the number of strong lenses observed will allow the inner structure (radial profile and shape) of dark matter halos in the lensing galaxies, groups of galaxies and clusters of galaxies to be studied with unprecedented accuracy.
\item The wide field-of-view will allow rare mergers of extremely massive clusters to be photographed and the depth and angular resolution will allow for a detailed mapping of dark matter matter in multiple merging clusters, perhaps rivaling the Bullet Cluster.  
\end{itemize} 
For some of the above goals LSST is better suited, while for others it is WFIRST. We refer the reader to the science cases for LSST and WFIRST~\cite{Abell:2009aa,Spergel:2013tha} for more details and references. 

Strong lensing is the most direct way to ``image" dark subhalos. A true measurement of the subhalo abundance in a galaxy using strong lensing will require a separate effort from future wide and deep galaxy surveys. Three promising methods, each with its own systematics but capable of measuring dark clumps with masses around $10^8 {\rm M}_\odot$ or lower, have been discussed recently. Presence of a small clump near an arc created by strong lensing of a background galaxy can distort the surface brightness distribution of the arc from being smooth (locally). This localized deviation from smoothness can be used to place constraints on the mass of the clump in lenses at cosmological distances~\cite{2012Natur.481..341V}. The time delays between multiple images created by gravitational lensing can be measured by monitoring sources like active galactic nuclei (AGN) that vary over time. These time delays are sensitive to substructure~\cite{Moustakas:2009na} and a mission OMEGA~\cite{Moustakas:2008ib} has been proposed to measure the substructure fraction in galaxies using time delays. A third method advocates using radio telescopes (like ALMA) to detect gravitational lensing of dusty, star forming galaxies bright in sub-mm wavelengths. With spectroscopy, it is expected that gravitational lensing of these bright, dusty, star forming galaxies is akin to having multiple sources (at the same redshift) being lensed by a foreground galaxy, increasing the signal to noise dramatically~\cite{Hezaveh:2012ai}. 
 
The lack of substructure, if any, should also be visible in the power spectrum measurements. In the past, this has been accomplished through the measurements of correlations in the Ly-$\alpha$ forest formed by absorption in neutral hydrogen clouds along the line of sight to quasars. It has been argued that the constraints from Ly-$\alpha$ preclude warm dark matter from solving any of the small-scale structure problems~\cite{Seljak:2006qw,Viel:2013fqw}. The main uncertainties in this method have to do with the unknown properties of the intergalactic medium~\cite{Viel:2012sd}. This is an issue that is systematics limited since the current sample of quasars is already huge~\cite{Palanque-Delabrouille:2013gaa} and will grow by an order of magnitude or more. In order to make progress, this method will require significant advances in the analysis methods and theory work including numerical simulations of the intergalactic medium. 

In comparison to warm dark matter, the primary effect of self-interactions of dark matter is to reduce the central
density of dark matter halos and create constant density (spherical)
cores.  Such effects would be expected to be observable if the cross section to particle mass
ratio is of order $1~\cm^2/\g$ (0.2-2 $\times 10^{-24}$ cm$^2$/GeV)~\cite{Spergel:1999mh,Vogelsberger:2012ku,Rocha:2012jg}. 
Cross sections of this magnitude can be produced in hidden sector dark matter models through the
exchange of a light gauge boson and this interaction can also endow
the dark matter particle with the desired relic density~\cite{Feng:2009hw}. 
In these cases, the Bullet Cluster constraints~\cite{Randall:2007ph} 
are typically weak because of the intrinsic velocity 
dependence of the cross section in these (and most other) models. 
This and other aspects of complementarity between direct searches for self-interacting dark matter 
and astrophysics are discussed further in section~\ref{sec:self}. 

In addition to structure formation, non-gravitational interactions of
dark matter could impact a variety of other astrophysical
phenomena. For example, coupling of axions and light sterile neutrinos (or
generally any light hidden-sector particles) to standard model
particles may affect the cooling of compact objects (stars, neutron
stars, white dwarfs, supernovae)~\cite{Raffelt:2000kp} or, decays or annihilations of dark matter particles could affect the process of reionization of neutral atoms, which took place in the first billion years after the Big Bang~\cite{Zhang:2007zzh}. 

While dark matter physics may have imprinted tell-tale astrophysical
signatures, it will be hard to unambiguously identify such signatures
as non-gravitational interactions of dark matter. The complementarity
with direct, indirect or collider searches is an essential part of
this endeavor.

%\section{Complementarity}
%\label{sec:complementarity}

\section{Qualitative Complementarity}
\label{sec:basic}

The various techniques and detection strategies described in the previous section are each able to provide different and complementary information about the nature dark matter.  These experimental approaches are each able to measure or constrain different aspects of the dark matter's interactions and other characteristics, and are in many cases sensitive to different classes of dark matter candidates. Furthermore, these different strategies are each limited by different astrophysical, instrumental, and other uncertainties. And while it may be unlikely than any one experimental endeavor will conclusively identify the particle nature of dark matter, a diverse array of experiments and observations could potentially bring together a collection of information over the coming decade that might bring resolution to the long-standing question of dark matter.

%As evident from the brief descriptions in \secref{probes}, every
%experimental approach provides useful information for every dark
%matter scenario.  At the same time, each approach is subject to
%different systematic uncertainties and no approach will illuminate all
%aspects of dark matter.  In detail, what is learned from each approach
%is highly scenario-dependent.

At a qualitative level, the complementarity between these categories of dark matter search strategies is illustrated by the
following observations that follow from the basic features of each approach:
\begin{itemize}
\setlength{\itemsep}{1pt}\setlength{\parskip}{0pt}\setlength{\parsep}{0pt}
\item {\em Direct Detection} is perhaps the most straightforward
  detection method, with excellent prospects for improved sensitivity
  in the coming decade and for discovering WIMPs.  The approach requires
  careful control of low-energy backgrounds, and is relatively
  insensitive to dark matter that couples to leptons only, or to
  WIMP-like dark matter with mass $\sim$$1~\gev$ or below. Ultimately, direct detection experiments could constrain the dark matter particle's mass, and its elastic scattering cross sections with protons and neutrons (modulo uncertainties in the local dark matter density).
  
\item {\em Indirect Detection} is potentially sensitive to dark matter interactions with all standard model particles, through combinations of gamma-ray, cosmic ray, neutrino, and multi-wavelength observations. These approaches are currently beginning to probe dark matter particles with annihilation cross sections similar to that predicted for a simple thermal relic, and experimental
  sensitivities are expected to improve significantly on several fronts within
  the coming decade.  Discovery through indirect detection requires
  understanding of astrophysical backgrounds, and prospects are often subject to uncertainties in the dark matter distribution. 
  
  \item {\em Particle Colliders}, such as the Large Hadron Collider, provide the opportunity to study dark
  matter in a highly controlled laboratory environment, and are potentially sensitive to a wide variety of dark matter models.  
  Hadron (lepton) colliders are relatively insensitive to dark matter that interacts
  only with leptons (hadrons). 
  Unlike direct and indirect astrophysical searches for dark matter, colliders will not be able to determine whether any newly discovered weakly interacting particle is stable and cosmologically relevant, or merely long-lived on the timescales relevant to its detectors ($\sim 100$ ns).

\item {\em Astrophysical Probes} can be sensitive to the ``warmth''
  of dark matter and to properties such as its
  self-interaction strength. By measuring the impact of the dark matter properties on the structure formation of the Universe, astrophysical probes may be able to identify departures from the cold, collisionless dark matter paradigm.  
  
  \end{itemize}

\section{Quantitative Complementarity}
\label{sec:quantitative}

\subsection{Effective Operator Description}

Many of the qualitative features outlined above can be illustrated in a simple
and fairly model-independent setting by considering dark matter that
interacts with standard model particles through four-particle contact
interactions, which represent the exchange of very heavy particles.
While not entirely general, these contact interactions are expected to work well to describe
theories in which the exchanged particle mass is considerably
larger than the momentum transfer of the physical process of interest.

In this toy exercise, we consider a spin-1/2 dark matter particle, $\chi$, 
with the following generation-independent interactions to quarks $q$, gluons $g$, 
and leptons $\ell$ (including neutrinos), respectively:
\begin{equation}
\frac{1}{M_q^2} ~\bar{\chi} \gamma^\mu\gamma_5 \chi \sum_q \bar{q} \gamma_\mu\gamma_5  q
+ \frac{\alpha_S}{M_g^3}~ \bar{\chi}  \chi G^{a \mu \nu} G^a_{\mu \nu}
+ \frac{1}{M_\ell^2} ~
\bar{\chi} \gamma^\mu  \chi \sum_\ell \bar{\ell} \gamma_\mu  \ell~.
\label{lagrangian}
\end{equation}
Although we could have chosen many other interactions (e.g., operators with different Lorentz structures,
or involving other SM particles), these three are reasonably representative examples which capture 
many of the phenomenological features most relevant for dark matter searches \cite{Goodman:2010yf}. 
For example, 
the interactions with quarks described above lead to spin-dependent elastic scattering with 
nuclei, while the interaction with gluons mediates a spin-independent interaction.  
The coefficients $M_q$, $M_g$, and $M_\ell$ each characterize the strength
of the interaction with the respective standard model particle.
%, and in this representative example should be chosen such that the combined
%annihilation cross section into all three channels provides the
%correct relic density of dark matter.  
The values of the three
interaction strengths, together with the mass of the dark matter
particle, $m_\chi$, completely define this theory (at sufficiently low energies) and allow one to
predict the rates of both spin-dependent and spin-independent direct
scattering, the annihilation cross sections into quarks, gluons, and
leptons, and the production rate of dark matter at 
colliders \cite{Beltran:2010ww,Goodman:2010ku,Bai:2010hh,Goodman:2010yf,Rajaraman:2011wf,Fox:2011pm,Bai:2012xg,Carpenter:2012rg}.

\begin{figure}[tb]
\includegraphics[width=0.32\columnwidth]{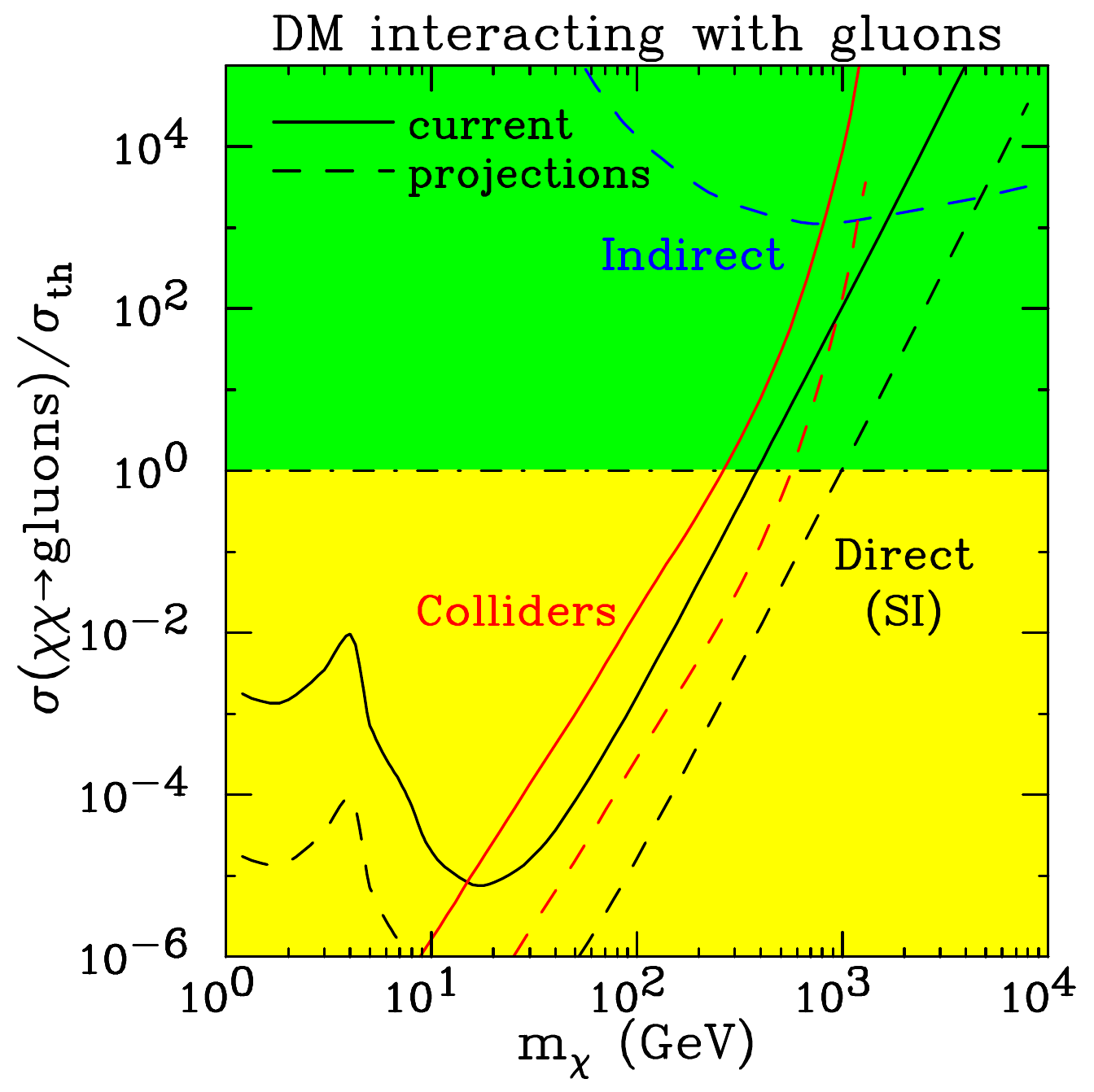}
\includegraphics[width=0.32\columnwidth]{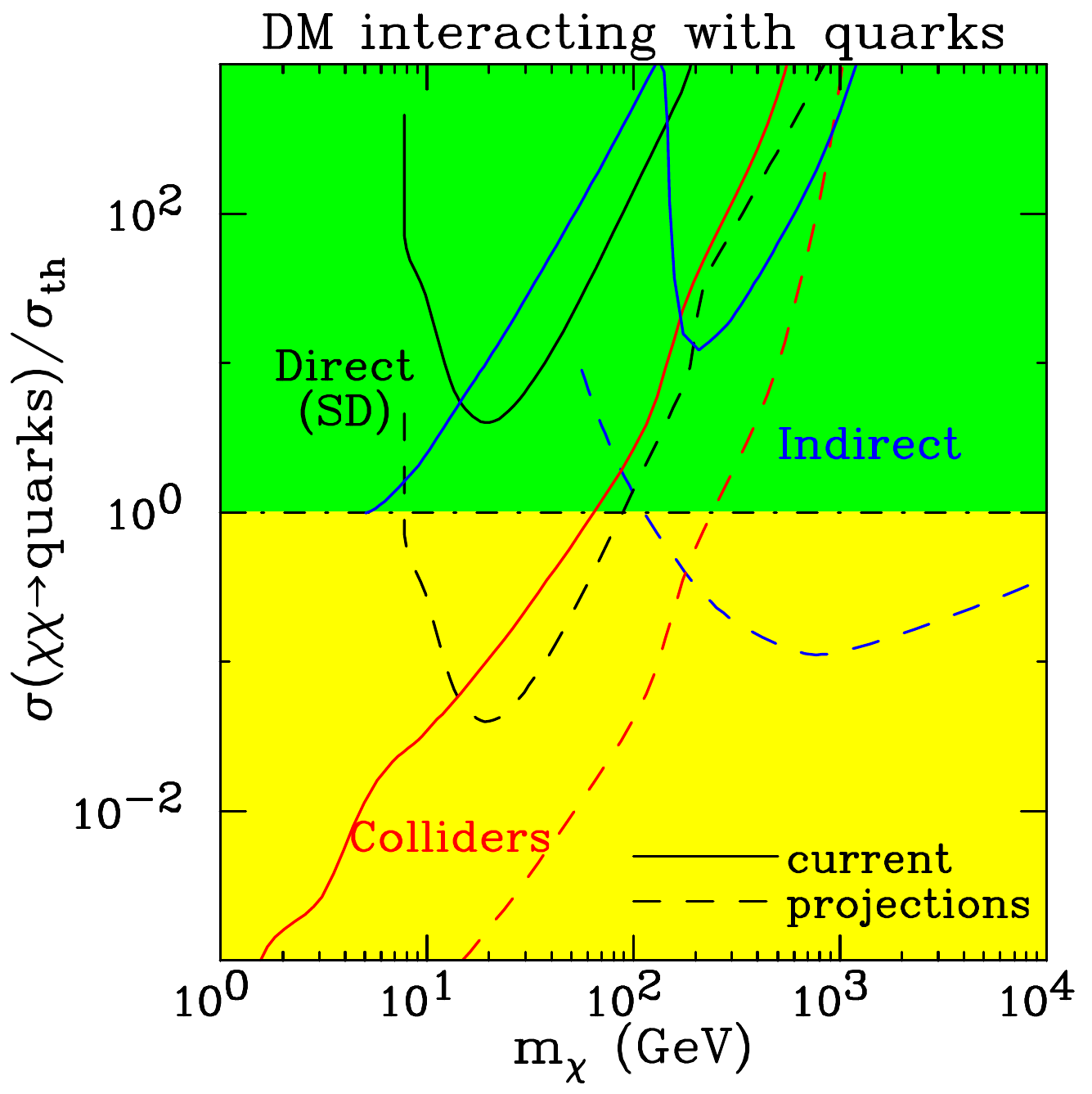}
\includegraphics[width=0.32\columnwidth]{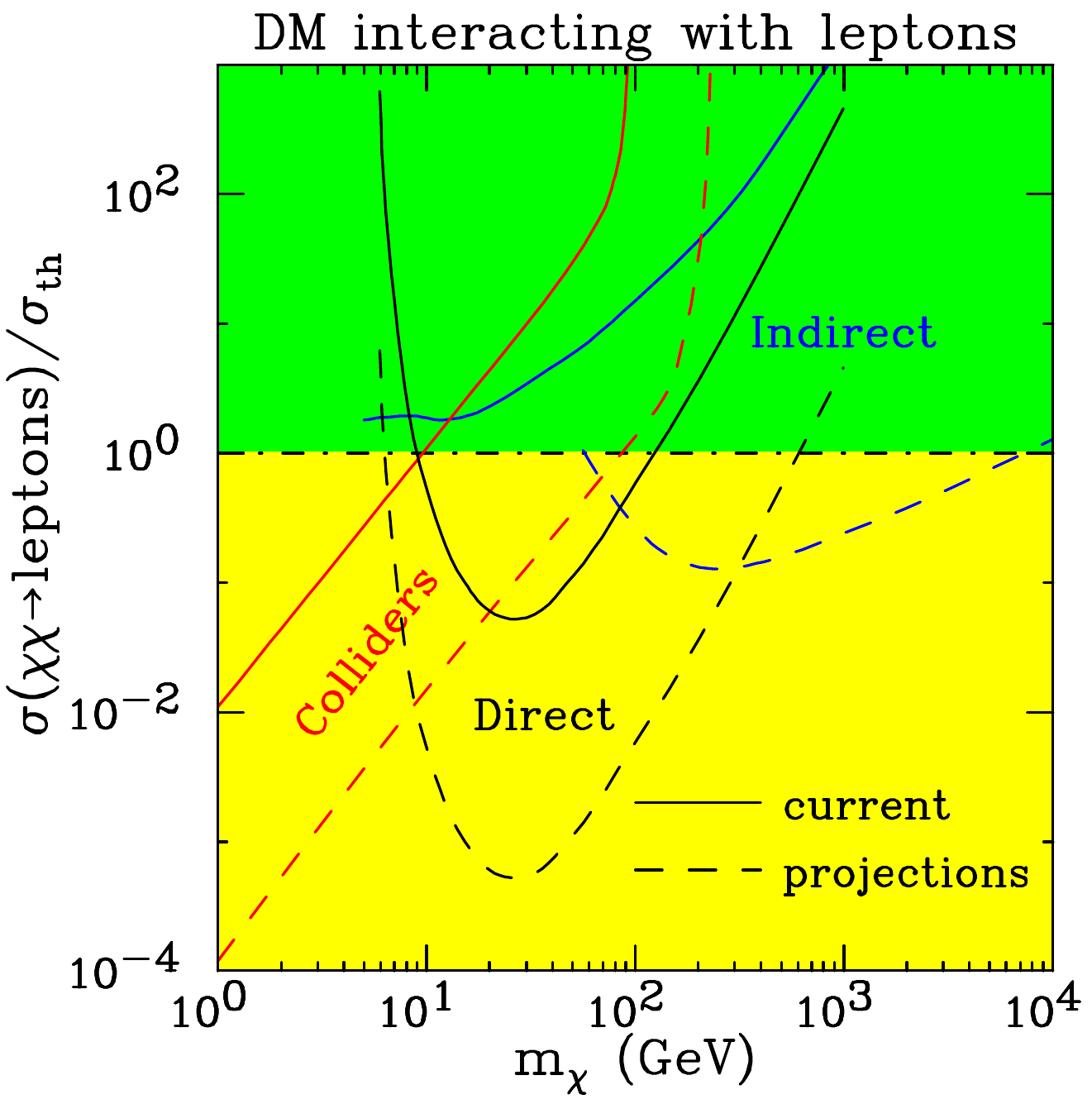}
%\hfil
%\includegraphics[width=0.49\columnwidth]{DMQuarkSDOps_sigma} \\
%\includegraphics[width=0.49\columnwidth]{DMLeptonicOps_sigma} 
%\vspace*{-.1in}
\caption{Dark matter discovery prospects in the $(m_{\chi},
  \sigma_i/\sigmath)$ plane for current and future direct
  detection~\cite{DMtools}, indirect
  detection~\cite{Ackermann:2011wa,IndirectCTA}, and particle
  colliders ~\cite{Chatrchyan:2012me,ATLAS:2012ky,Vacavant:2001sd} for
  dark matter coupling to gluons~\cite{Goodman:2010ku},
  quarks~\cite{Goodman:2010ku,Beltran:2008xg}, and
  leptons~\cite{Fox:2011fx,Chae:2012bq}, as indicated.
\label{fig:prospects}}
\end{figure}

Each class of dark matter search outlined in \secref{probes} is
sensitive to some range of the interaction strengths for a given dark
matter mass.  Therefore, they are all implicitly capable of putting a bound on that interaction's contribution to the annihilation cross section into a particular channel.  Since the annihilation cross section at the temperature of freeze-out predicts the dark matter relic density, the
reach of any experiment is thus equivalent to a fraction of the
observed dark matter density $\Omega_{\rm DM}$.  This connection can be seen in the
plots in \figref{prospects}, which show the annihilation cross sections
$\sigma_i(M_i), (i=g,q,l)$ for each individual channel,
normalized to the value $\sigmath$ required\footnote{For 
non-thermal WIMPs or asymmetric dark matter, the annihilation cross-section
does not have a naturally preferred value, but the plots in \figref{prospects}
are still meaningful.} for a single thermal WIMP
\beq
\frac{\sigma_i(M_i)}{\sigmath} = \frac{f_i}{\Omega_\chi/\Omega_{\rm DM}},
\label{Omegascaling}
\eeq
where
\beq
f_i \equiv \frac{\sigma_i(M_i)}{\sigma_{\rm total}} \le 1
\eeq
is the fractional contribution of the particular channel $i$ to the total annihilation 
cross-section $\sigma_{\rm total}\equiv \sigma(\chi\chi\to anything)$.

Assuming $f_i=1$ (i.e., that an individual channel $i$ saturates the total annihilation 
cross-section $\sigma_{\rm total}$), $\sigmath$ provides a natural target for 
dark matter searches: if the discovery potential for an experiment reaches 
cross sections $\sigma_i\sim \sigmath$ (the horizontal dot-dashed lines in 
\figref{prospects}), that experiment will be able to discover a thermal relic 
which could potentially (with the assumption of $f_i=1$) account
for all dark matter in the universe. On the other hand, 
if an experiment were to observe an interaction consistent with an
annihilation cross section $\sigma_i$ below $\sigmath$ (yellow-shaded regions in
\figref{prospects}), it would have discovered dark matter, but through an 
interaction that cannot alone account for the observed relic abundance.
In that situation, $\Omega_\chi\sim \OmegaDM$ can be achieved only with $f_i<1$
and therefore, additional interactions contributing to other annihilation channels
in addition to $i$ must have been present in the early universe.  
Finally, if an experiment were to observe a cross section $\sigma_i$
above $\sigmath$ (green-shaded regions in \figref{prospects}), this discovery 
could point to a multicomponent dark matter scenario, in which the $\chi$ 
species is only one among several dark matter particles ($\Omega_\chi < \OmegaDM$).
Alternatively, it can also be suggestive of dark matter with a non-thermal origin
(in which case eq.~(\ref{Omegascaling}) does not apply), 
or of dark matter with interactions that are 
not well-described by the effective operator approach of eq.~(\ref{lagrangian}).
Further discussion can be found in Refs.~\cite{Profumo:2013hqa,Frandsen:2012rk,An:2012va,Bai:2013iqa,An:2013xka,Chang:2013oia,DiFranzo:2013vra,Busoni:2013lha,Buchmueller:2013dya}.

In \figref{prospects}, we assemble the current bounds (solid lines)
and discovery potential (dashed lines) for several near-term dark 
matter searches that are sensitive to interactions with gluons, quarks, or leptons.  
It is clear that the searches are complementary to each other in terms of being
sensitive to interactions with different standard model particles.
These results also illustrate that within a given interaction type,
the reach of different search strategies depends sensitively on the
dark matter mass.  For example, direct searches for dark matter are
very powerful for masses around 50 GeV (even in the lepton scenario, 
due to loop-mediated $\gamma$ and $Z$ exchanges \cite{Fox:2011fx}),
but have difficulty at very low masses, where the dark matter particles carry 
too little momentum to noticeably affect heavy nuclei. This region of low mass is
precisely where collider production of dark matter is expected to be most promising, 
since high energy collisions readily produce light dark matter particles 
with large momenta.  \figref{prospects} confirms that the difficult low-mass region is effectively 
covered by searches at the LHC (in the case of gluon or quark couplings)
or at LEP and ILC (in the case of lepton couplings).
The sensitivity of both direct searches and colliders is increasingly diminished
at high masses, and this is where indirect detection probes
play an important complementary role --- in the case of couplings to quarks and leptons,
CTA arrays are able to cover the relevant parameter region in the mass range around $1$ TeV.

\subsection{Supersymmetry}

The effective theory description of the dark matter
interactions with standard model particles discussed above is an attempt to capture
the salient features of the dark matter phenomenology without
reference to any specific theoretical model.  However, the
complementarity between the different dark matter probes illustrated in
\figref{prospects} is also found when one considers specific theoretical models with WIMP 
dark matter candidates.  Among the many other alternatives,
low energy supersymmetry~\cite{Martin:1997ns} and models with universal extra dimensions (UED) 
have been a popular and widely studied extensions of the standard model. 
In the following, we will discuss the interplay between different dark matter detection 
strategies within the context of supersymmetric and UED models. 

%As a concrete example, we will consider the lightest neutralino, $\tilde\chi^0_1$, as our dark matter candidate, within the context of the minimal supersymmetric standard model.

%\subsubsection{Neutralino Dark Matter in Various Supersymmetric Frameworks}

\subsubsection{The phenomenological MSSM (pMSSM)}

% and we
%shall use it here as our second example. In supersymmetry, the
%dark matter candidate is generally the lightest neutralino $\tilde\chi^0_1$,
%which is its own anti-particle.

Even within the general context of low-energy supersymmetry, there are many
different scenarios that can be considered. Philosophically, these scenarios tend 
to fall within either top-down or bottom-up approaches. In top-down models, 
the low-energy sparticle spectrum is derived from a high-energy theory, which 
relies on various theoretical assumptions, but typically requires relatively few 
input parameters. Alternatively, one can phenomenologically describe the properties 
of the low-energy sparticle spectrum with fewer theoretical assumptions, 
but with a greater number of input parameters.  We begin this discussion of 
complementarity within the context of supersymmetric dark matter by considering 
a phenomenological approach to the minimal supersymmetric standard model 
(MSSM), based on the results presented in \cite{Cahill-Rowley:2013dpa},
before turning our attention to more concrete theoretical scenarios later on. 

\begin{table}
\centering
\begin{tabular}{|c|c|} \hline\hline
$m_{\tilde L(e)_{1,2,3}}$ & $100 \gev - 4 \tev$ \\ 
$m_{\tilde Q(q)_{1,2}}$ & $400 \gev - 4 \tev$ \\ 
$m_{\tilde Q(q)_{3}}$ &  $200 \gev - 4 \tev$ \\
$|M_1|$ & $50 \gev - 4 \tev$ \\
$|M_2|$ & $100 \gev - 4 \tev$ \\
$|\mu|$ & $100 \gev - 4 \tev$ \\ 
$M_3$ & $400 \gev - 4 \tev$ \\ 
$|A_{t,b,\tau}|$ & $0 \gev - 4 \tev$ \\ 
$M_A$ & $100 \gev - 4 \tev$ \\ 
$\tan \beta$ & 1 - 60 \\
$m_{3/2}$ & 1 eV$ - 1 \tev$ ($\tilde{G}$ LSP)\\
\hline\hline
\end{tabular}
\caption{Scan ranges for the 19 (20) parameters of the pMSSM with a neutralino (gravitino) LSP. The gravitino mass is scanned with a log prior. 
All other parameters are scanned with flat priors, though this choice is expected to have little qualitative impact on the results~\cite{Berger:2008cq,Conley:2010du,Conley:2011nn}.}
\label{ScanRanges}
\end{table}

In the pMSSM approach, one scans over all phenomenologically relevant input parameters 
and considers all models which pass the existing experimental constraints and have a dark matter
candidate which can account for at least a portion of the observed dark
matter density~\cite{CahillRowley:2012cb,CahillRowley:2012rv,CahillRowley:2012kx}.  
The pMSSM parameters and the ranges of values employed in the scans are listed in Table~\ref{ScanRanges},
where the lower and upper limits were chosen to be essentially consistent with Tevatron and LEP data 
and to have kinematically accessible sparticles at the LHC, respectively. To study the pMSSM, 
many millions of model points were generated in this space (using SOFTSUSY~\cite{Allanach:2001kg} 
and checking for consistency with SuSpect~\cite{Djouadi:2002ze}, while 
the decay patterns of the SUSY partners and the extended Higgs sector 
are calculated using a modified version of SUSY-HIT~\cite{Djouadi:2006bz}). 
These individual models are then subjected to a large set of collider, flavor, precision measurement, 
dark matter and theoretical constraints~\cite{CahillRowley:2012cb}.  

\begin{figure}[t]
\includegraphics[width=0.47\columnwidth]{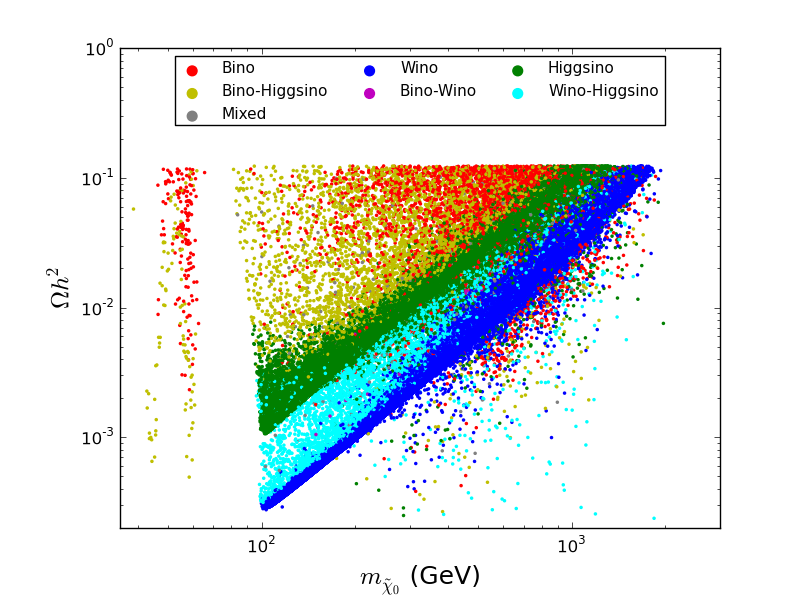}
\includegraphics[width=0.47\columnwidth]{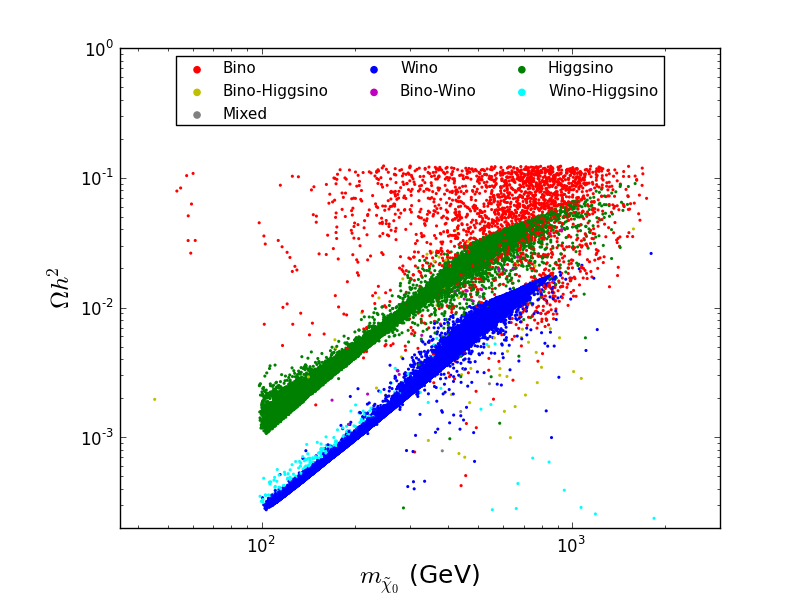}
%\vspace*{-0.10cm}
\caption{Left: Thermal relic density as a function of the LSP mass in the pMSSM model set, as generated, color-coded by the electroweak properties of the  LSP as discussed in the text. Right: Thermal relic density as a function of the LSP mass for all pMSSM models, surviving after all searches, color-coded by the electroweak properties of the LSP.}
\label{fig00}
\end{figure}

Roughly 225k models with a neutralino LSP 
survive this initial selection and can then be used for further physics studies. 
%Results from such model-independent scans with over 200,000 points are
%shown in \figref{pMSSM}, where each dot represents one particular
%supersymmetric model. 
The left panel in Figure~\ref{fig00} shows the thermal relic densities of the LSPs in the pMSSM model sample 
as a function of their mass, with suitable color-coding reflecting their electroweak eigenstate content
as indicated in the figure. Figure~\ref{fig00} illustrates essentially every possible mechanism to 
obtain (or lie below) the required relic density $\OmegaDM$: ($i$) The set of points at low masses 
forming ``columns" corresponds to bino and bino-Higgsino admixtures surviving due to their proximity 
to the $Z$ and $h$-funnels, where the annihilation cross-section is resonantly enhanced. ($ii$) 
The gold-colored points saturating the relic density in the upper left represent models with bino-Higgsino 
LSPs of the so-called ``well-tempered" variety. ($iii$) the red pure\footnote {Here `pure' means having an 
eigenstate fraction $\geq 90\%$.} bino models in the middle top of the Figure are bino co-annihilators (mostly with 
sleptons) or  are models near the $A$-funnel region. ($iv$) The green (blue) bands are pure 
Higgsino (wino) models that saturate the relic density bound near $\sim 1(1.7)$ TeV but dominantly appear at far lower relic densities. Wino-Higgsino hybrids are seen to lie between these two cases as expected. 
%($v$) A smattering of other annihilator models are seen to be loosely distributed in the lower right-hand corner of the figure. 

\begin{figure}[t]
\includegraphics[width=0.65\columnwidth]{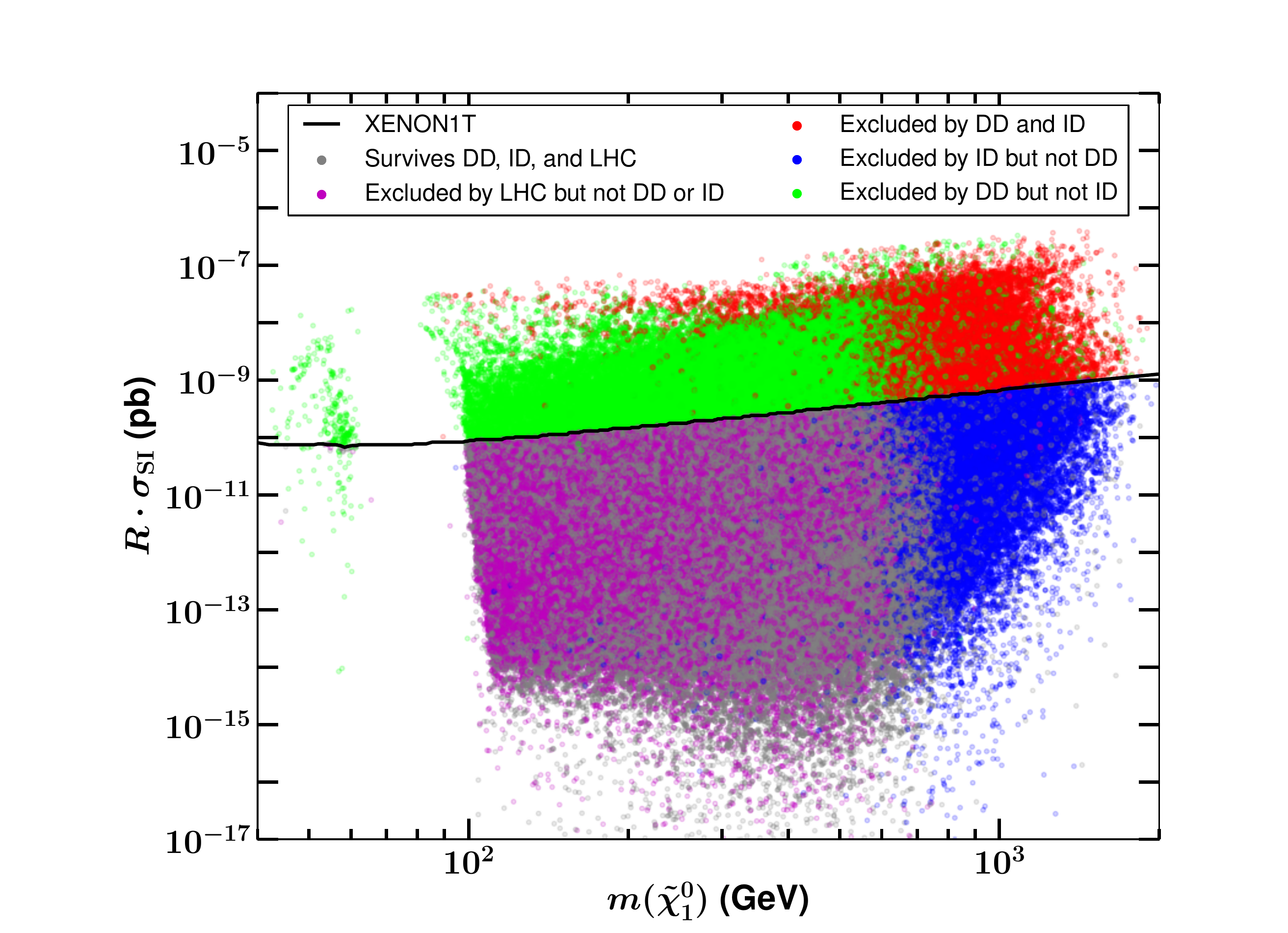}
\caption{Results from a model-independent
  scan~\cite{CahillRowley:2012cb,Cahill-Rowley:2013dpa}
%,CahillRowley:2012rv,CahillRowley:2012kx} 
of the parameter space in the minimal supersymmetric model
(MSSM), presented in the $(m_\chi,R\cdot \sigma_{\text{SI}}^p)$ plane.
%or the $(m_\chi,m_{\text{LCSP}})$ plane (right panel). 
The models are divided into categories, depending on whether
dark matter can be discovered in future direct detection experiments
(green points), indirect detection experiments (blue points) or both
(red points). Among the remaining models, the magenta (grey) points 
represent models that will (will not) be discovered at 
the upgraded LHC.
\label{fig:pMSSM}}
\end{figure}

Within each model, the dark matter interactions are completely specified 
and one can readily compute all relevant dark matter signals.
Fig.~\ref{fig:pMSSM} shows the survival and exclusion rates resulting from the 
various searches and their combinations in the plane of the LSP mass versus 
the spin-independent direct detection cross section, scaled by a factor 
$R=\Omega_\chi/\OmegaDM$ to account for the fact that most of the 
pMSSM models lead to a thermal relic density somewhat below the measured 
value $\OmegaDM$.
The models are the categorized depending on the
observability of a dark matter signal in direct detection
experiments (green points), indirect detection experiments (blue points)
or both (red points). The left panel in Fig.~\ref{fig:pMSSM} contrasts 
direct detection (DD) and indirect detection (ID) searches.
Note that the DD- and ID-excluded regions are relatively well separated in 
terms of mass and cross section although there is also some overlap between 
the sets of models excluded by the different experiments. 
This nicely illustrates the complementarity between DD and ID probes:
at low and moderate masses, the DD experiments cover a large fraction 
of models (green points) which would otherwise evade indirect detection.
At the same time, at high masses, a sizable fraction of models
(the blue points) can only be seen in indirect detection (via ground-based gamma ray telescopes).  
The red points represent models for which one would have an independent confirmation 
of a dark matter signal by two different types of probes. 
Among the models which escape detection in DD and ID dark matter experiments,
some  (the magenta points in Fig.~\ref{fig:pMSSM}) are already ruled out by current LHC data
while others (the remaining grey points) will potentially be probed after the LHC upgrade 
(the analysis of the projected sensitivity of the upgraded LHC is still ongoing). 
\Figref{pMSSM} demonstrates that the three different
dark matter probes nicely combine to discover most (albeit not all)
supersymmetry models in this scan. 

The models which survive all searches are
categorized in the right panel of Fig.~\ref{fig00}, which should be compared 
with the original model set as generated in the left panel of Fig.~\ref{fig00}. 
One can see that ($i$) the models in the light $h$ and $Z$-funnel 
regions have essentially evaporated (and further measurements of the invisible 
width of Higgs as well as the standard SUSY searches will likely further restrict these), 
($ii$) the well-tempered neutralinos are now seen to be completely gone; 
($iii$) the possibility of almost pure Higgsino or 
wino LSPs even approximately saturating the relic density has vanished thanks 
to CTA, ($iv)$ the mixed wino-Higgsino models, due to a combination of measurements, 
have also completely disappeared. ($v$) The only models remaining which 
{\it do} saturate the WMAP/Planck relic density are those with bino co-annihilation.

\subsubsection{Specific MSSM scenarios}

One of the driving motivations for supersymmetry (SUSY) theories is
the hierarchy problem of the Standard Model (SM).  The presence of
superpartners at the weak scale cancels quadratic divergences in the
Higgs potential, drastically reducing the fine-tuning present in the
theory.  However, current LHC results generically imply squark masses
of $\gtrsim 1~\tev$.  In most SUSY models this implies a mild reintroduction
of fine-tuning, which, while far less severe than the fine-tuning
present in the SM, weakens the motivation of SUSY theories.

This issue is compounded by the discovery of the Higgs boson at the
LHC with a mass of $m_h \approx
125.6~\gev$~\cite{Aad:2012tfa,Chatrchyan:2012ufa}.  In the MSSM, 
it is well-known that radiative corrections can lift the Higgs mass above the 
tree-level bound of $m_Z \simeq 91~\gev$.  However,
achieving a Higgs mass as large as the one observed experimentally
requires either stop masses of $\mathcal{O}(8-10)~\tev$ or a large
stop $A$-term.  Large $A$-terms are non-generic, while stop masses of
$\mathcal{O}(8-10)~\tev$ seem to reintroduce some fine-tuning.

Relatively heavy scalars are also motivated from precision observables
in the flavor and CP violation sectors.  General weak-scale SUSY
models suffer serious constraints from flavor violating observables,
which would point to a heavy sfermion sector.  
Even though well-known mechanisms exist to ameliorate the dangerous
flavor-changing effects, in concrete models, CP violation is generally still present, 
and motivates sfermion masses in the multi-TeV range to avoid electron and
neutron EDM constraints~\cite{Feng:2000bp}.

{\bf Focus Point Supersymmetry.}
One specific framework which addresses this combination of issues is focus
point (FP) supersymmetry~\cite{Feng:1999mn,Feng:1999zg}.  In the MSSM
for $\tan\beta \gtrsim 5$, electroweak symmetry breaking requires at tree level
\begin{equation}
m_Z^2 \approx - 2 \mu^2 - 2 \mhu^2 (\mweak) \ ,
\label{mz2}
\end{equation}
where $\mu$ is the higgsino mass parameter and $m_{H_u}^2$ is the soft 
up-type Higgs mass parameter.  The theory is relatively natural, 
if $m_Z^2 \sim \mu^2 \approx \left| \mhu^2 \right|$,
%while it becomes fine-tuned when
%$m_Z^2 \ll \mu^2 \approx \left| \mhu^2 \right|$.  The former condition
%condition is of course satisfied if all SUSY-breaking masses are
%$\mathcal{O}(M_{\mathrm{weak}})$, but it can also be satisfied if
%SUSY-breaking masses are significantly larger than $m_Z$ at the
%SUSY-breaking scale but $\mhu^2 \rightarrow 0$ at the weak scale.
which can be achieved for certain sets of boundary conditions due to renormalization group (RG) running, 
even when (some of) the SUSY-breaking masses are significantly larger than $m_Z$.
%This mechanism is present in FP SUSY models, where the Higgs potential
%exhibits ``radiative naturalness'' due to renormalization group (RG)
%running.
One such example is the FP scenario, in which 
%The required boundary conditions are dependent on the SUSY-breaking
%scale, and for $M_{\mathrm{SUSY-breaking}} \sim M_{\mathrm{GUT}}$ one
%solution is the case of unified 
the scalar masses are universal at the GUT scale, the $A$-terms are negligible (see, however \cite{Feng:2012jfa}), and
the gaugino masses are small.  This scenario is realized in the constrained
MSSM (CMSSM) with $A_0 = 0$ and $M_{1/2} \ll m_0$, producing the
``focus point region'' of the CMSSM where $m_0$ is in the multi-TeV
range but $\mu$ remains near the weak scale \cite{Feng:1999mn,Feng:1999zg}.  

While the collider prospects for the FP region are poor due to the large
superpartner masses, large portions can be probed at a variety of dark
matter experiments.  
%Neutralino dark matter is perhaps the most
%studied dark matter candidate, and it is well-known that the simplest
%cases produce a thermal relic density either larger or smaller than
%the observed dark matter abundance.  A pure Bino is generally
%overabundant while a pure Wino or Higgsino is underabundant for
%neutralino masses $m_\chi\lesssim1~\tev$, and some further mechanism
%is required to bring the relic density in agreement with the observed
%dark matter abundance.  
Given the relatively small values of $\mu$, the FP region exhibits a 
mixed Bino-Higgsino LSP, which produces the right amount of dark matter.  
Significant Bino-Higgsino
mixing also enhances both direct detection and annihilation signals,
improving the ability of dark matter experiments to probe FP models
\cite{Feng:2000gh,Feng:2000zu}.
Spin-independent direct detection is particularly interesting for such
models, with cross-sections within the reach of current and
near-future direct detection experiments~\cite{Feng:2010ef}.

\begin{figure}[tb]
  \subfigure[$m_0$ and $m_\chi$ for $\Omegachi \approx 0.23$.]{
    \includegraphics[width=0.48\columnwidth]{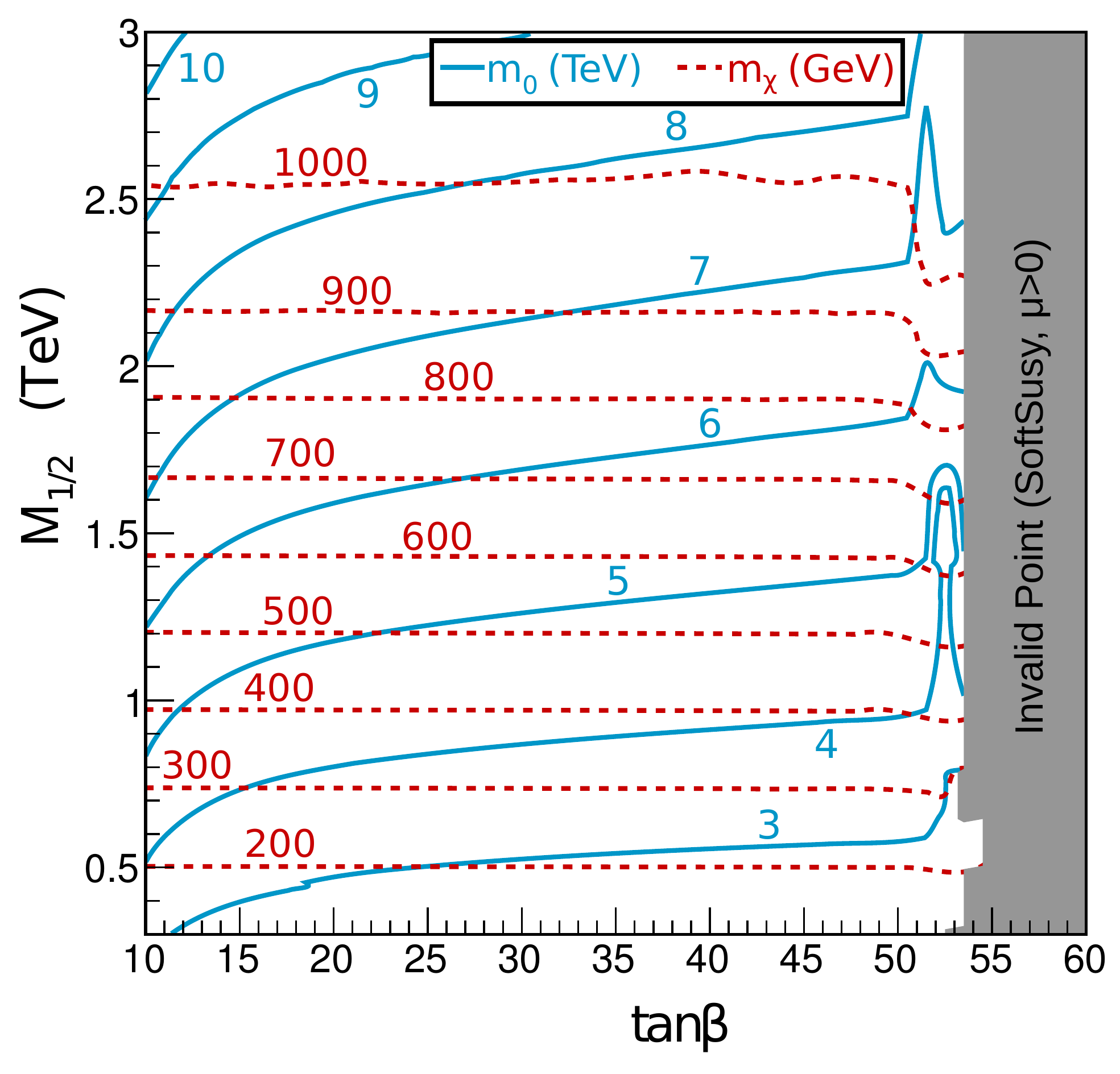}
    \label{fig:m0mchi} }
  \subfigure[Limits on FP SUSY for $\Omegachi \approx 0.23$.]{
    \includegraphics[width=.48\textwidth]{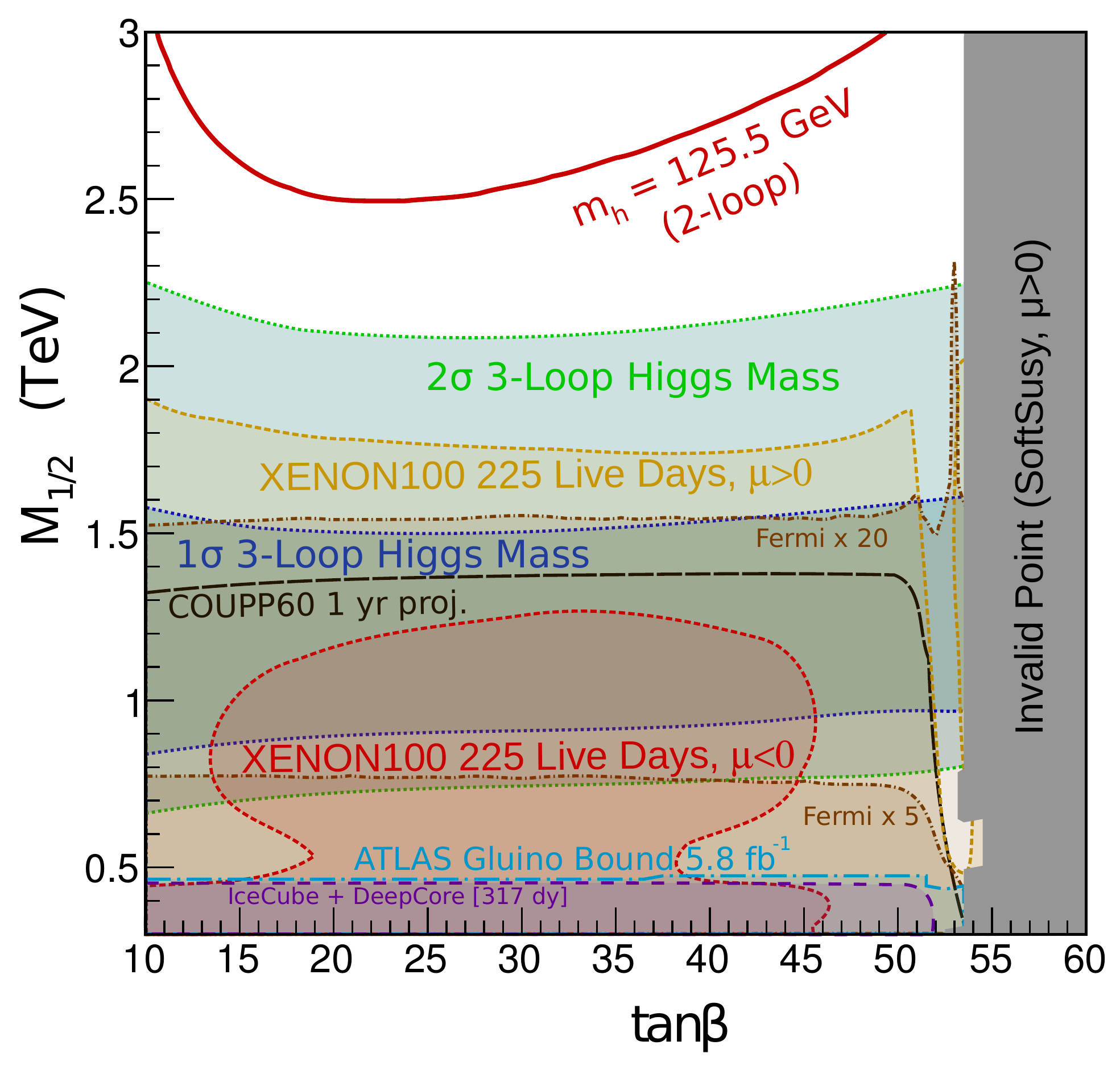}
    \label{fig:alllimits} }
\vspace*{-.1in}
\caption{\label{fig:fp} \textit{The Focus Point Region for $\Omegachi
    = \OmegaDM$.}  Shown are the values of $m_0$ and $m_\chi$
  satisfying the thermal relic density (left) and relevant constraints
  (right).  The collider constraint is extrapolated from ATLAS gluino
  searches using $5.8 fb^{-1}$ of data with
  $\sqrt{s}=8~\tev$~\cite{ATLAS:2012ona}.  Current dark matter
  constraints use results from XENON100 with 225 live
  days~\cite{Aprile:2012nq} and IceCube (w/ DeepCore) results using
  317 live days~\cite{Aartsen:2012kia}.  Projected limits are drawn from
  projected COUPP60 sensitivity after a 1 year physics
  run~\cite{Ramberg:2010zz} and multiples of the current Fermi-LAT
  sensitivity from a stacked dwarf spheroidal analysis~\cite{Atwood:2009ez}.
  The two-loop Higgs mass is determined by
  \textsc{SOFTSUSY}~\cite{Allanach:2001kg}, while the three-loop
  result uses \textsc{H3m}~\cite{Harlander:2008ju,Kant:2010tf}.  Plots
  are taken from Ref.~\cite{Draper:2013cka}.}
\end{figure}

In exploring the FP parameter space, a particularly interesting region
is the slice wherein the neutralino thermal relic density saturates
the observed relic density,
\begin{equation}
\Omegachi\left( m_0, M_{1/2}, A_0, \tan\beta,
\mathrm{sign}\left(\mu\right) \right) = \OmegaDM\, .
\label{OmegaFP}
\end{equation}
By fixing the sign of $\mu$ and setting $A_0 = 0$ to generate the
desireable FP RG behavior, for a particular choice of $\left\{M_{1/2},
\tan\beta \right\}$ there exists a unique value of $m_0$ for which
$\Omegachi = \OmegaDM$~\cite{Feng:2011aa}.  This allows the FP region
to be studied in the $\left\{M_{1/2}, \tan\beta \right\}$ plane
wherein \textit{every point has the correct thermal relic density} \cite{Feng:2011aa}.
By directly taking into account the cosmological constraint 
(\ref{OmegaFP}), one is able to better explore a larger portion of the relevant 
parameter space.
%Constraints on the FP region using this framework were studied in
%Refs.~\cite{Feng:2011aa,Draper:2013cka}, the latter of which (and
%plots here
Results for the FP region are shown in \Figref{fp}, based on the relic density
determination of WMAP with 7 years of data~\cite{Komatsu:2010fb}.
The analysis also used \textsc{SOFTSUSY}~3.1.7~\cite{Allanach:2001kg} for
spectrum generation, \textsc{MicrOMEGAs}~2.4~\cite{Belanger:2010gh} to
calculate the relic density and direct detection processes, and
\textsc{DarkSUSY}~5.0.5~\cite{Gondolo:2004sc} to calculate indirect detection
rates.  \Figref{m0mchi} shows the value of $m_0$ required to achieve
$\Omegachi = \OmegaDM$ and the corresponding neutralino mass $m_\chi$.
Generally the required value of $m_0$ increases with increasing
$M_{1/2}$ and decreases with increasing $\tan\beta$.  However, for
$\tan\beta \gtrsim 50$ the appropriate value shifts downward, as the
pseudoscalar Higgs bosons becomes light enough in this region to
significantly alter the relic density calculation.

\Figref{alllimits} shows associated constraints upon the FP region.
Due to the relatively high squark mass scale, the dominant collider constraints
are derived from searches for gluino
pair-production~\cite{ATLAS:2012ona}.  The associated reach is
relatively limited, constraining $M_{1/2} \lesssim 500~\gev$ almost
independent of $\tan\beta$.  Even with current data, direct detection
experiments place significantly stronger bounds, with current
XENON100~\cite{Aprile:2012nq} results constraining $M_{1/2} \gtrsim
1.8~\tev$ for nearly the entire range of $\tan\beta$ shown if $\mu >
0$, with somewhat stronger bounds at large and small $\tan\beta$.  For
$\mu < 0$ the constraints are weaker, requiring $M_{1/2} \gtrsim
1~\tev$ for a moderate range of $\tan\beta$ but weakening to $M_{1/2}
\gtrsim 500~\gev$ for small $\tan\beta$ and placing no constraint for
$\tan\beta \gtrsim 45$.  Spin-independent limits were produced using a
strange quark form factor of $f_s =
0.05$~\cite{Freeman:2009pu,Young:2009zb,Giedt:2009mr}.  While current
spin-dependent results do not have sensitivity to the focus point
parameter space, near future results from COUPP60~\cite{Ramberg:2010zz} are
expected to constrain $M_{1/2} \lesssim 1.3-1.4~\tev$ for $\tan\beta
\lesssim 50$.  Indirect detection experiments are also relevant, with
current IceCube~\cite{Aartsen:2012kia} results constraining $M_{1/2} \gtrsim
500~\gev$ for $\tan\beta \lesssim 50$, producing a bound competitive
with current LHC constraints.  Moreover, while gamma ray searches
currently do not probe the FP region, an order of magnitude
improvement on current Fermi-LAT sensitivity from dwarf
spheroidals~\cite{Atwood:2009ez} will provide significant sensitivity to
$M_{1/2} \sim 700~\gev - 1.5~\tev$.  The generation of IceCube and
Fermi-LAT bounds are detailed in Ref.~\cite{Draper:2013cka}.  Future
experiments on all these fronts will have the ability to probe
significantly larger regions of FP parameter space.

These sensitivities are especially important given recent computations
of the Higgs mass with leading 3-loop effects included in models with
multi-TeV scalars~\cite{Draper:2013cka,Feng:2013tvd}.  Generally
2-loop determinations require stop masses of $\mathcal{O}(8-10)~\tev$,
which in the FP region with $A_0 = 0$ places the appropriate Higgs
mass at $M_{1/2} \gtrsim 2.5~\tev$, beyond the reach of current or
near-future experiments.  However, including 3-loop effects reduces
the required stop mass to $3-4~\tev$ even without left-right mixing,
improving the prospect of dark matter and collider experiments to
probe FP models, with a favored region of $700~\gev \lesssim M_{1/2}
\lesssim 2.2~\tev$~\cite{Draper:2013cka}.

\begin{figure}[tbp]
\begin{center}
\includegraphics[width=0.47\textwidth]{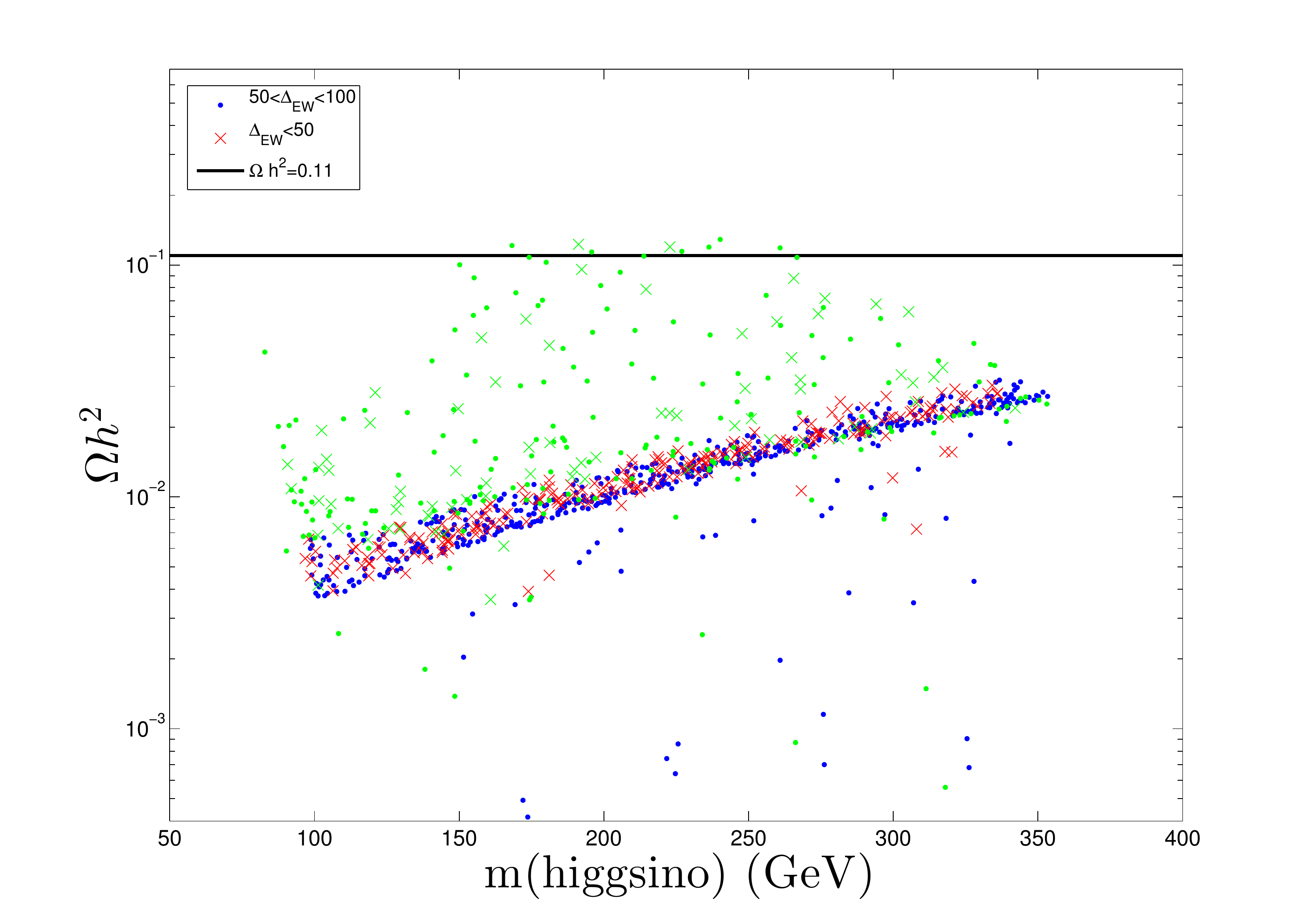}
\includegraphics[width=0.47\textwidth]{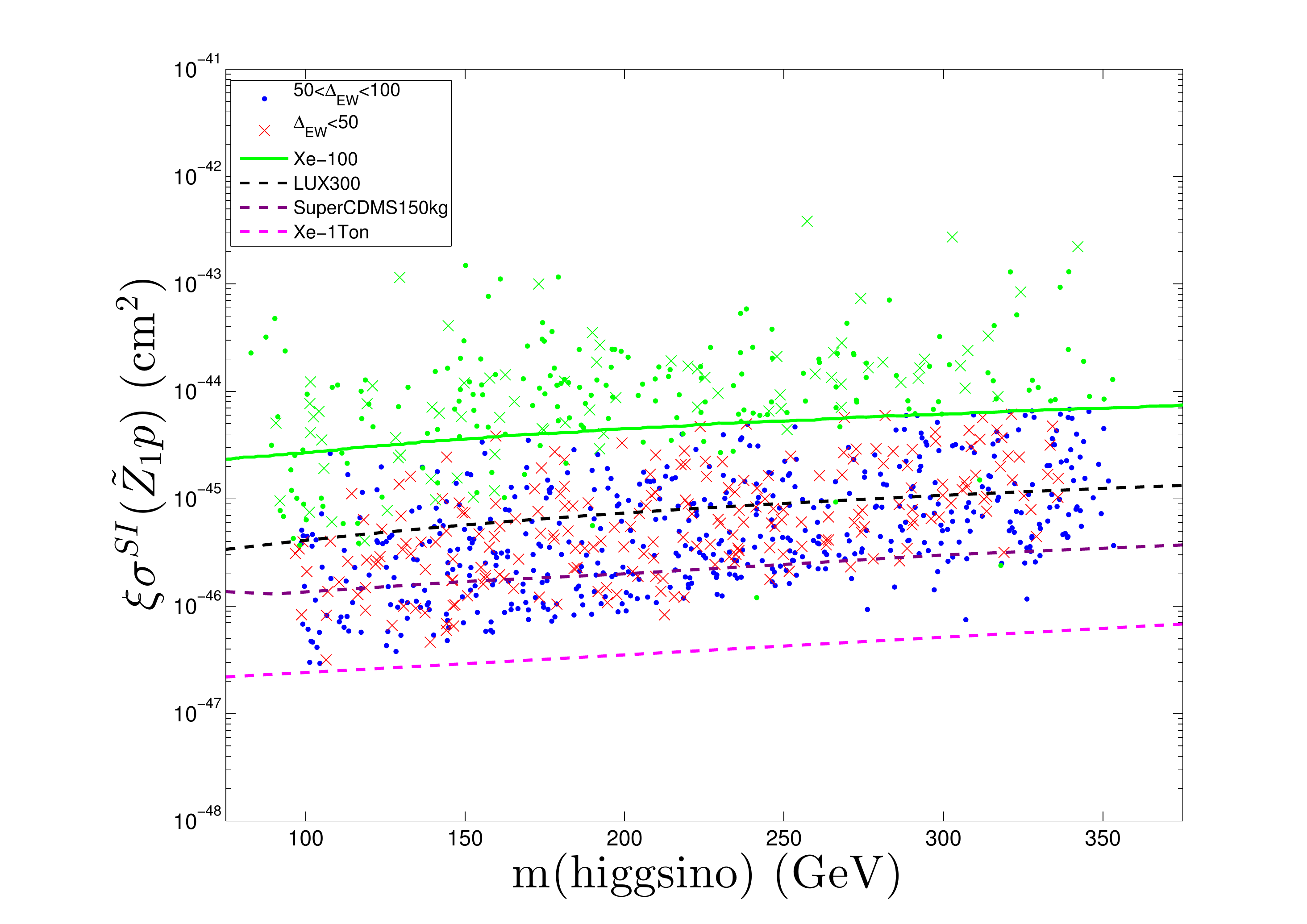}
\caption{Left: Plot of the standard thermal neutralino abundance 
$\Omega_{\tz_1}^{TP}h^2$ versus $m(higgsino)$ 
from a scan over NUHM2 parameter space with $\Delta_{EW}<50$ (red crosses) and $\Delta_{EW}<100$ (blue dots). 
Green points are excluded by current direct/indirect WIMP search experiments.
The horizontal line shows the central value of $\Omega_{CDM}h^2$ from WMAP9.
Right: Plot of rescaled higgsino-like WIMP spin-independent
direct detection rate $\xi \sigma^{SI}(\tz_1 p)$
versus $m(higgsino)$ from a scan over NUHM2 parameter space with $\Delta_{EW}<50$ (red crosses)
and $\Delta_{EW}<100$ (blue dots).
Green points are excluded by current direct/indirect WIMP search experiments.
Also shown are the current limits from XENON100 experiment,
and projected reaches of LUX, SuperCDMS 150 kg and XENON1T.
\label{fig:oh2}}
\end{center}
\end{figure}

{\bf Radiatively-driven Natural Supersymmetry.}
Another way to ameliorate the fine tuning present in eq.~(\ref{mz2}) 
is to consider non-universal boundary conditions for the soft SUSY 
breaking parameters at the GUT scale.
%In previous studies \cite{Baer:2012up,Baer:2012cf,Baer:2013ava} it has been argued that 
%a necessary condition for naturalness of SUSY models is the requirement of no
%large uncorrelated cancellations to $m_Z^2/2$ in the one-loop effective potential
%minimization condition
%%
%\be \frac{m_Z^2}{2} =
%\frac{m_{H_d}^2 + \Sigma_d^d -
%(m_{H_u}^2+\Sigma_u^u)\tan^2\beta}{\tan^2\beta -1} -\mu^2 \;.
%\label{eq:loopmin}
%\ee 
%%
%Here, Eq.~(\ref{eq:loopmin}) is implemented as a {\it weak scale
%relation}, even for SUSY theories purporting to be valid all the way up
%to scales as high as $M_{\rm GUT}-M_{P}$.  The quantities $\Sigma_u^u$
%and $\Sigma_d^d$ are the one-loop corrections arising from loops of
%particles and their superpartners that couple directly to the Higgs
%doublets. Thus, electroweak naturalness requires 1. low $\mu\sim
%100-300$ GeV, 2. $m_{H_u}^2$ is driven to small negative values so that
%$m_{H_u}^2({\rm weak})\sim 100-300$ GeV and 3. highly mixed TeV-scale top
%squarks. The large mixing both reduces the $\tst_1$ and $\tst_2$
%contributions to $\Sigma_u^u$ while lifting the Higgs mass $m_h\sim 125$
%GeV. 
One specific such class of models resulting in a low value of
the fine-tuning measure $\Delta_{EW}$ has been labeled as
radiatively-driven natural supersymmetry (RNS) \cite{Baer:2012up,Baer:2012cf,Baer:2013ava}
(see also \cite{Cohen:1996vb,Falk:1999py,Bagger:1999ty,Ellis:2002wv}).
In RNS models, $\mu\sim 100-300$~GeV, and the lightest neutralino
is largely higgsino-like, albeit with some non-negligible gaugino component.
%\footnote{ In the RNS model, if gaugino mass $M_3$ is too
%large, it lifts the top squark masses high enough so that the radiative
%corrections $\Sigma_u^u(\tst_{1,2})$ become large leading to
%fine-tuning.  Since the gaugino masses $M_1$ and $M_2$ are related to
%$M_3$ under gaugino mass unification, then the lightest neutralino--
%while dominantly higgsino-like-- always has a non-negligible gaugino
%component. In models with light higgsinos but with very large gaugino masses
%(such as Br\"ummer-Buchm\"uller model\cite{Brummer:2012zc}), the lightest neutralino
%is nearly pure higgsino.  In that case, the spin-independent direct
%detection rates presented below {\it will not obtain} since the
%$h-\tz_1\tz_1$ coupling depends on a product of higgsino and gaugino
%components\cite{Baer:2013vpa}.}  
The thermally-produced relic density $\Omega_{\tilde h}^{TP}$ of
higgsino-like WIMPs from RNS is typically found to be a factor $5-15$ below the
measured value for CDM,  as shown in Fig. \ref{fig:oh2}~\cite{Baer:2012cf,Baer:2013vpa}.  
To accommodate this
situation, a cosmology with mixed axion/higgsino dark matter (two dark
matter particles, an axion and a higgsino-like neutralino)\cite{Choi:2008zq,Baer:2011hx,Baer:2011uz,Bae:2013qr}
has been invoked in Ref.~\cite{Baer:2012cf,Baer:2013vpa}.  In this case, thermal
production of axinos $\ta$ in the early universe followed by $\ta\to
g\tg,\ \gamma\tz_i$ leads to additional neutralino production. In the
case where axinos are sufficiently produced, their decays may lead to
neutralino re-annihilation at temperatures below freeze-out; the
resulting re-annihilation abundance is always larger than the standard
freeze-out value.  In addition, coherent-oscillation production of
saxions $s$ at high PQ scale $f_a>10^{12}$~GeV followed by saxion decays
to SUSY particles can also augment the neutralino abundance. Late saxion
decay to primarily SM particles can result in entropy dilution of all
relics (including axions) present at the time of decay, so long as BBN
and dark radiation constraints are respected.  The upshot is that,
depending on the additional PQ parameters, either higgsino-like
neutralinos or axions can dominate the dark matter abundance, or they
may co-exist with comparable abundances: this leads to the possibility
of detecting both an axion and a WIMP (see also Section~\ref{sec:postdiscovery} below).

In the case of mixed axion-WIMP dark matter, the local WIMP abundance
might be well below the commonly accepted local abundance
$\rho_{loc}\sim 0.3$~GeV/cm$^3$. Thus, to be conservative, limits from
experiments like XENON100 or CDMS should be compared to theoretical
predictions which have been scaled down\cite{Bottino:2000jx} by a factor
$\xi\equiv \Omega_{\tilde h}^{TP}h^2/0.12$.  In the RNS model, the
gauginos cannot be too heavy so that the neutralino always has a
substantial gaugino component even though it is primarily higgsino.
This means that spin-independent direct detection rates
$\sigma_{SI}(\tz_1p)$ are never too small.  Predictions for
spin-independent higgsino-proton scattering cross section are shown in
Fig. \ref{fig:oh2} and compared against current limits and future reach
projections.  
%                                                                              
%\begin{figure}[tbp]
%\begin{center}
%\includegraphics[height=0.3\textheight]{sc_omh2}
%\caption{Plot of rescaled higgsino-like WIMP spin-independent
%direct detection rate $\xi \sigma^{SI}(\tz_1 p)$
%versus $m(higgsino)$ from a scan over NUHM2 parameter space with $\Delta_{EW}<50$ (red crosses)
%and $\Delta_{EW}<100$ (blue dots).
%Green points are excluded by current direct/indirect WIMP search experiments.
%We also show the current limit from $Xe$-100 experiment,
%and projected reaches of LUX, SuperCDMS 150 kg and $Xe$-1 ton.
%\label{fig:SI}}
%\end{center}
%\end{figure}
%                                                                              
Even accounting for the local scaling factor $\xi$, it is
found \cite{Baer:2013vpa} that ton-scale noble liquid detectors such as
XENON1T \cite{Aprile:2012zx} should completely probe the model parameter
space. One caveat is that if saxions give rise to huge entropy dilution
after freeze-out while avoiding constraints from dark radiation and BBN,
then the local abundance may be even lower than the assumed freeze-out
value, and the dark matter would be highly axion dominated.

\subsubsection{The Next-to-minimal MSSM (NMSSM)}

Despite a great deal of attention, the MSSM is not without shortcomings, e.g.~the so-called 
``mu-problem'' is related to the presence of a term in the superpotential
\be
W_{MSSM}\supset \mu H_{u}H_{d}
\label{muterm}
\ee
with a mass parameter $\mu$ which has no a-priori connection to the electroweak scale
The Next-to-Minimal Supersymmetric Standard Model (NMSSM) is a simple extension to 
the MSSM which tries to alleviate the mu-problem by promoting the $\mu$ parameter 
to a dynamical field $S$, replacing (\ref{muterm}) with
\be
W_{NMSSM}\supset \lambda_{S}SH_{u}H_{d} + \frac{\kappa}{3}S^{3},
\ee
where $\lambda_{S}$ and $\kappa$ are dimensionless parameters.  When the singlet $S$ 
obtains a VEV, an effective $\mu$ parameter is generated.  The additional scalar superfield $S$ 
in the NMSSM 
%and similar singlet extended SUSY models 
increases the field content, leading to another possible dark matter candidate, 
the singlino, $\tilde S$~\cite{Barger:2005hb,Barger:2007nv,Djouadi:2008uw}.

\begin{figure}[t]
\begin{center}
     \includegraphics[angle=0,width=0.32\textwidth]{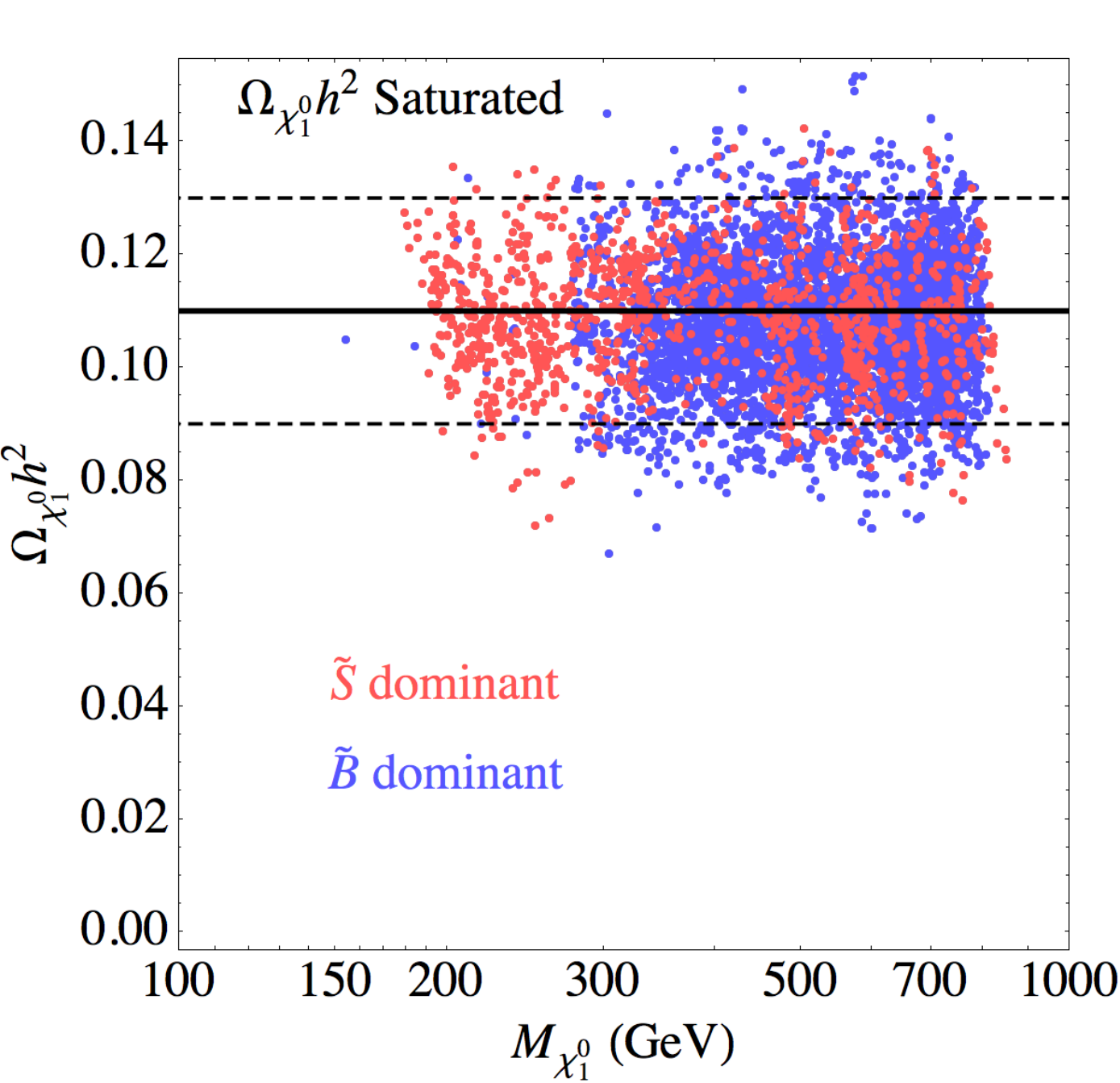}
     \includegraphics[angle=0,width=0.32\textwidth]{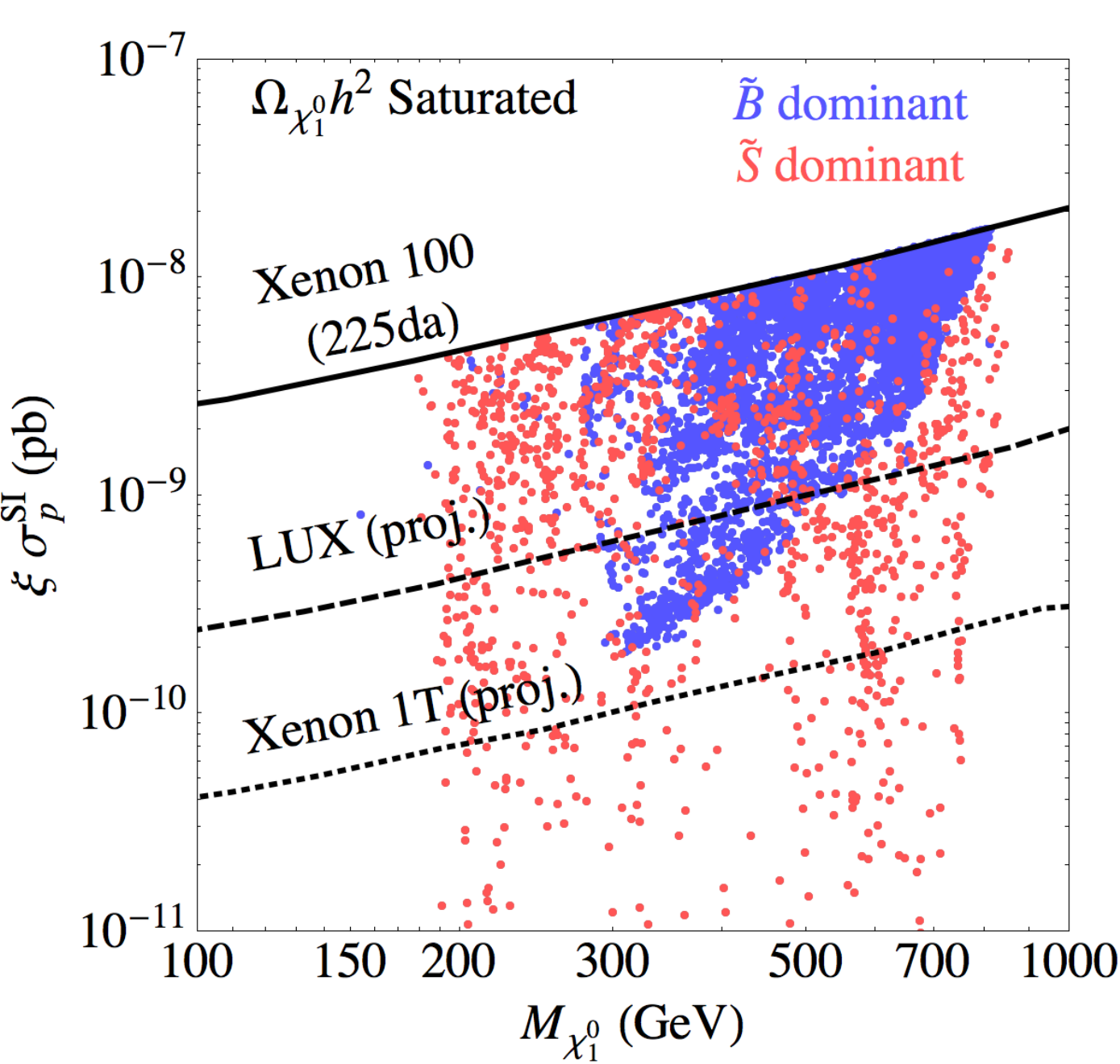}
     \includegraphics[angle=0,width=0.32\textwidth]{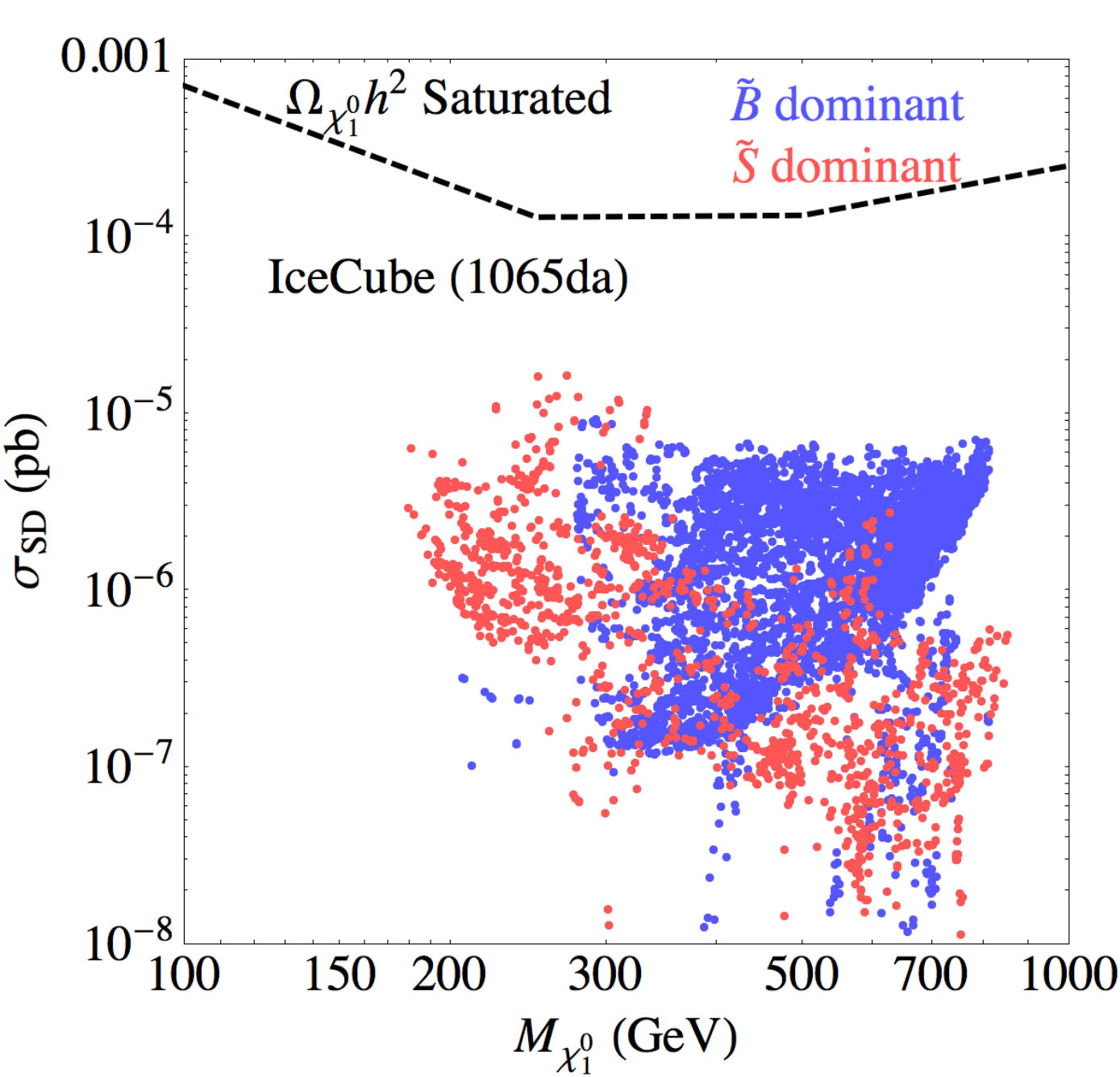}\\
     \includegraphics[angle=0,width=0.32\textwidth]{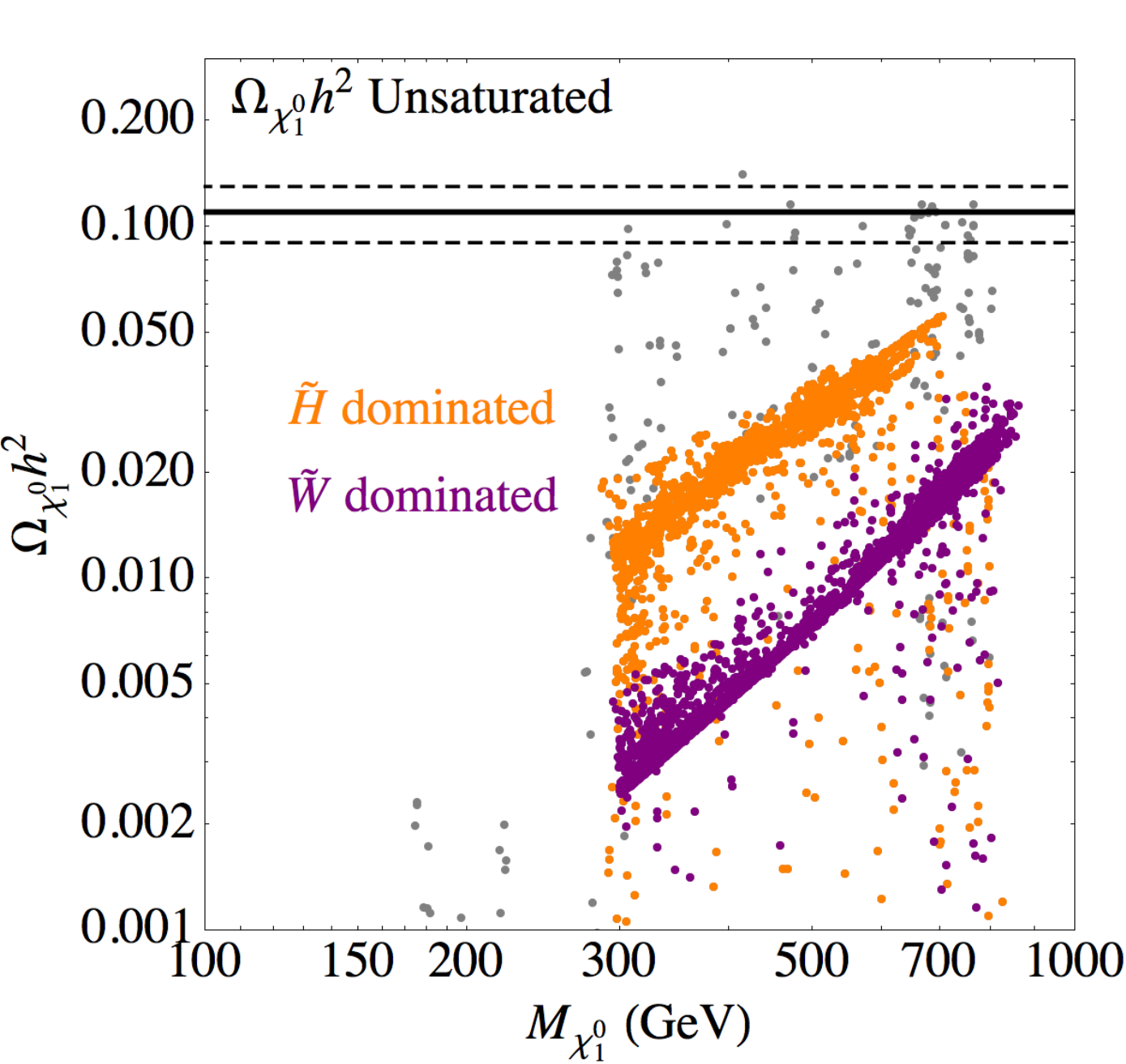}
     \includegraphics[angle=0,width=0.32\textwidth]{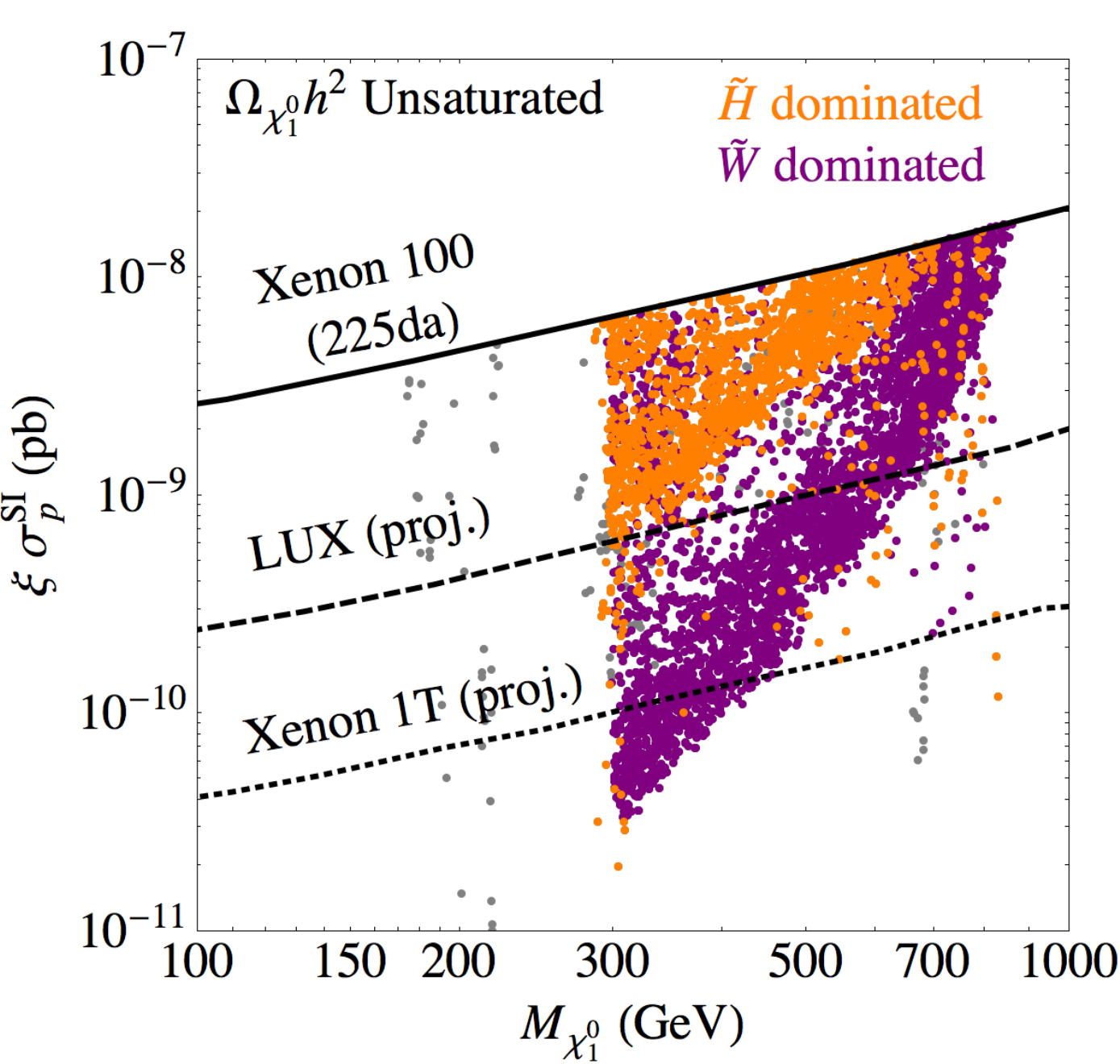}
     \includegraphics[angle=0,width=0.32\textwidth]{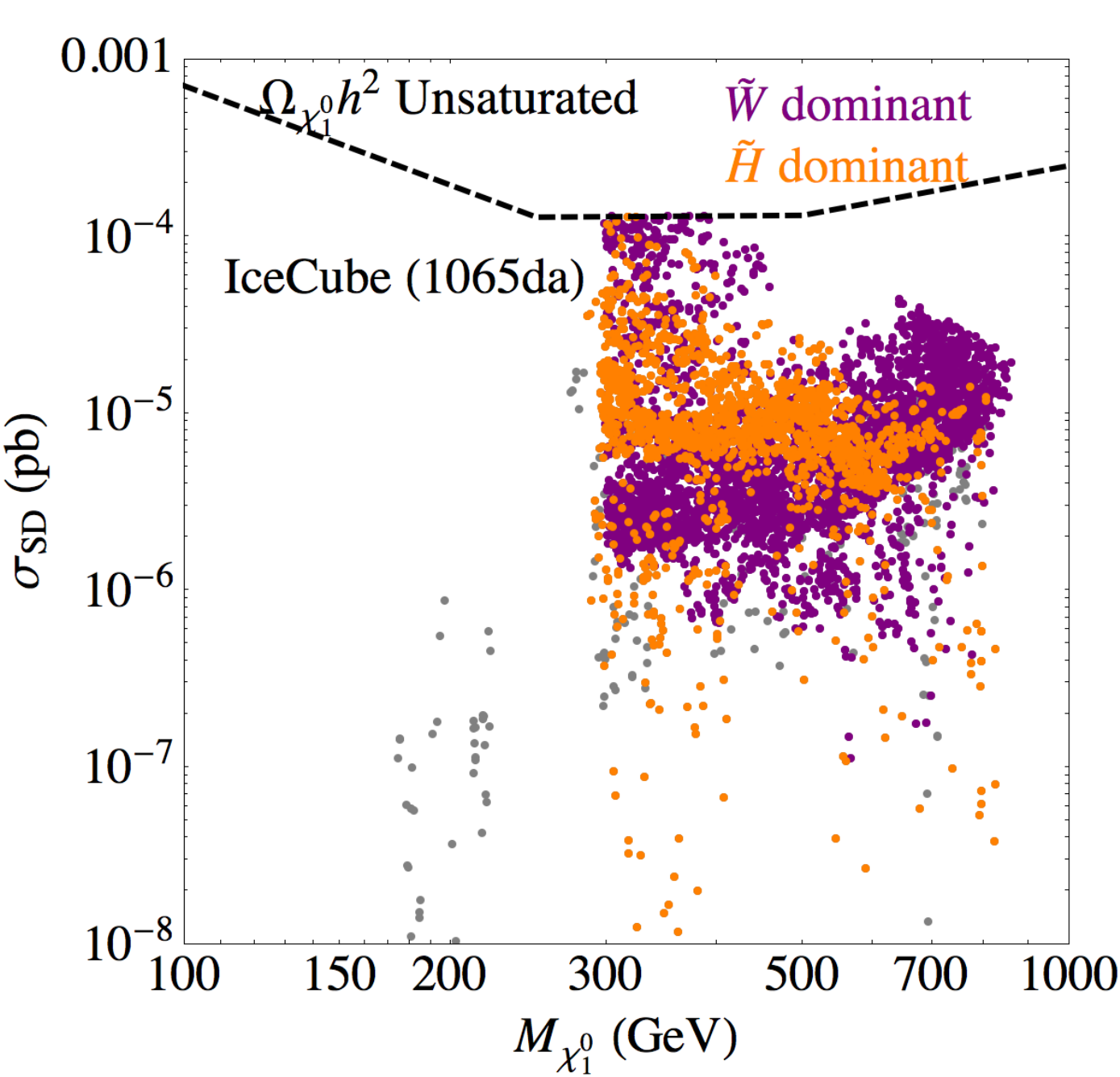}
\caption{Relic abundance (left), scaled SI cross section (middle) and SD cross section (right) for the saturated (top) and unsaturated (bottom) DM relic abundance scenarios.  Limits on $\sigma_{\rm SI}$ from Xenon100~\cite{Aprile:2012nq}, LUX~\cite{Akerib:2012ak}, and Xenon 1 Ton~\cite{Aprile:2012zx}, and $\sigma_{\rm SD}$ from IceCube~\cite{IceCube:2011aj}. Figures taken from Ref.~\cite{McCaskey:2013uka}.
  }
\label{figs}
\end{center}
\end{figure}

Here the status of the NMSSM with respect to current data is investigated, 
under the assumption of decoupled sfermions, in order to focus on the 
role of the singlet scalar and its superpartner, the singlino.  Two cases are explored,
first, where $\n$ can account for the dark matter relic abundance in its entirety, 
and second, when it is only a subdominant component.  The fit requires consistency 
with SUSY searches~\cite{Ellwanger:2004xm,Belanger:2008sj,ATLAS:2013rla} and 
present $h$(125) measurements at the LHC~\cite{ATLAS:2013sla,CMS:yva}. 
In the case of a saturated relic abundance, a 10\% theoretical uncertainty on the 
measured value of $\Omega_{DM}h^2$ is assumed. The relic abundance for 
the two cases are shown in the left panels of Fig.~\ref{figs}.  

%%%%%%%%%%%%%%%%%%%%%%%%%%%%%%%%%%%%%%%%%%%%%%%%
%%%%%%%%%%%%%%%%%%%%%%%%%%%%%%%%%%%%%%%%%%%%%%%%

The LHC $h$(125) measurements indicate a SM-like Higgs boson.  Therefore, if $\tan \beta \gg 1$, the lightest Higgs is typically dominated by $H_u$.  As a consequence, the $h \n\n$ coupling that primarily controls $\sigma_{\rm SI}$ avoids tension with the Xenon100 exclusion if $\n$ is mainly bino or wino.  When $\n$ is a wino or higgsino, the masses $\chi_2^0$ and $\c$ are comparable to $\n$, leading to significant co-annihilation effects.  Moreover, this scenario offers a difficult scenario at the LHC for discovering these states as the decay products will typically be soft and evade the selection cuts (recall the right panel in Fig.~\ref{fig00}).   

In the saturated relic abundance scenario, $\n$ can be either bino or singlino.  Some key observables are illustrated in the top row of Fig.~\ref{figs}.
\bi
\item In the bino scenario, $\sigma v_{\rm tot}$ is often below $3\times 10^{-27}$ cm$^3$ s$^{-1}$, therefore $\n\c$ and $\n\chi_2^0$ co-annihilation is required, with mass splitting of ${\cal O }(20\text{ GeV})$.  The SI scattering cross sections are within reach of future detectors such as Xenon 1 Ton and LUX. 
\item In the singlino scenario, $\sigma_{\rm SI}$ is reduced since usually dominant $h$ exchange is suppressed by the lack of a singlet component.  %The singlino case can have a generally smaller $\sigma_{\rm SI}$.
\ei
In either case, the SD cross section is more than a factor of 10 below the current bound placed by COUPP~\cite{Behnke:2012ys} and IceCube~\cite{IceCube:2011aj}.  A light scalar singlet with $M_s < 100$ GeV is possible, which may offer interesting signatures at the LHC such as $h\to ss\to b\bar b + b\bar b (\tau\tau)$.

The unsaturated case (lower row of plots in Fig.~\ref{figs}) is dominated by either higgsino or wino DM.  In both cases, co-annihilation with the associated $\c$ results in a low relic abundance, thus the direct detection sensitivities are scaled by the local density $\rho_{\rm \n} \propto \xi \equiv {\Omega_{\n} h^2/0.11}$.
\bi
\item The higgsino scenario accounts for the upper band of relic density and scaled SI cross section in the correspoding panels in Fig.~\ref{figs}, and is therefore within reach of future detectors such as LUX or Xenon 1 Ton.  
\item A wino dominated $\n$ is more suppressed than the higgsino scenario, and resides in the lower band of the relic abundance and $\sigma_{\rm SI}$ and may fall beyond the reach of Xenon 1 Ton.

\ei
Since the primary exchange for $\sigma_{\rm SD}$ occurs through a $Z$ boson, the cross section is proportional to the higgsino asymmetry.  This asymmetry is generally suppressed for large neutralino masses.  The current experimental sensitivity of COUPP is about an order of magnitude larger than the scaled scattering rate, $\xi \sigma_{\rm SD}$.  However, the sensitivity from IceCube does not critically depend on $\xi$.  Therefore, for masses below 500 GeV, IceCube may soon be sensitive to both higgsino and wino scenarios. 

\subsection{Universal Extra Dimensions}

In this section, we present updated results \cite{Arrenberg:2008wy,Arrenberg:2013paa}
on the complementarity between high-energy colliders and dark matter direct detection experiments 
in the context of Universal Extra Dimensions \cite{Appelquist:2000nn}.
As our reference, we take the mass spectrum in Minimal Universal Extra Dimensions (MUED), 
which is fixed by the radius ($R$) of the extra dimension and the cut-off scale 
($\Lambda$) \cite{Cheng:2002ab,Cheng:2002iz}.
To illustrate the complementary between dark matter detection and searches at the LHC, 
we introduce a slope in the MUED mass spectrum, in terms of the mass splitting ($\Delta_{q_1}$) between 
the mass of the lightest Kaluza-Klein (KK) partner (LKP) $m_{LKP}$ and the KK quark mass $m_{q_1}$,
$\Delta_{q_1} = \frac{m_{q_1} - m_{LKP}}{m_{LKP}}$. 
We take $\Delta_{q_1}$ as a free parameter, which is possible in a more general framework 
with boundary terms and bulk masses (see, e.g., \cite{Flacke:2013pla}).
The LKP is taken to be either the KK mode $\gamma_1$ of the photon (as in MUED),
or the KK mode $Z_1$ of the $Z$-boson. In the latter case, we assume the gluon and the 
remaining particles to be respectively 20\% and 10\% heavier than the $Z_1$. 
This choice is only made for definiteness, and does not impact our results, 
as long as the remaining particles are sufficiently heavy and do not participate in co-annihilation processes.

\begin{figure}[t]
\includegraphics[width=0.48\textwidth]{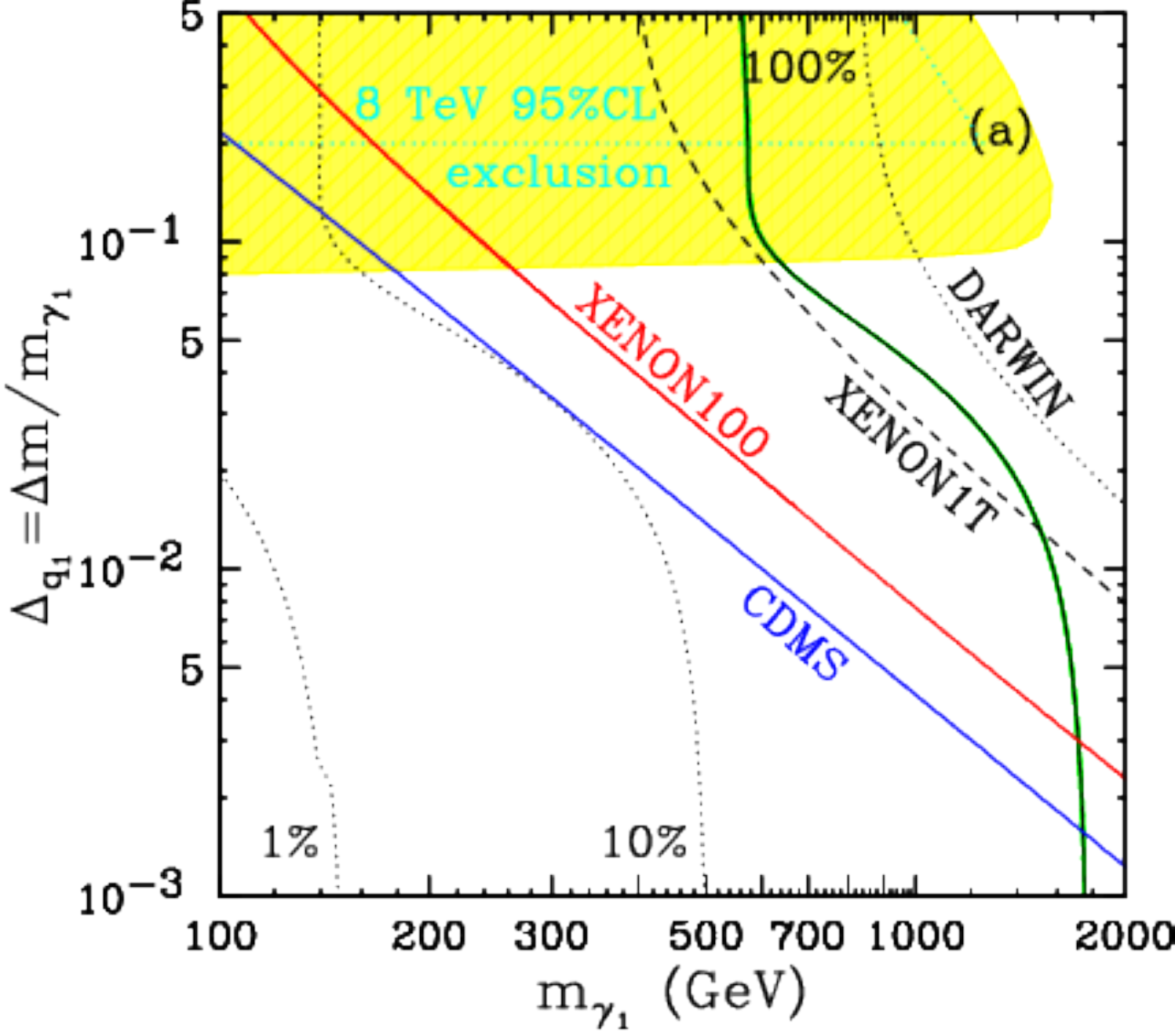}
\hspace{0.1cm}
\includegraphics[width=0.48\textwidth]{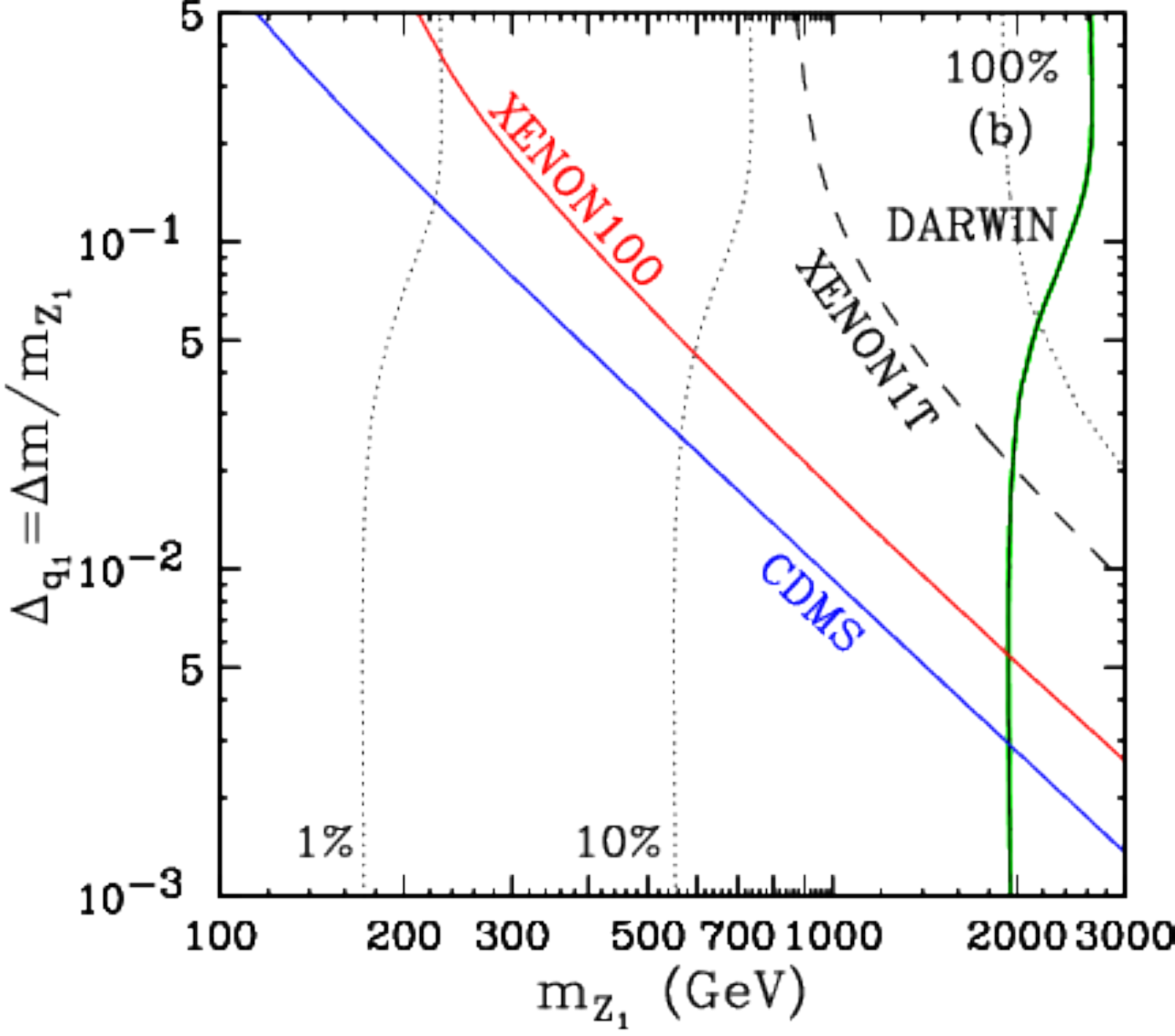}
\caption{\sl
Combined plot of the direct detection limit on the spin-independent cross section, 
the limit from the relic abundance and the LHC reach for (a) $\gamma_1$ and (b) $Z_1$, 
in the parameter plane of the LKP mass and the mass splitting $\Delta_{q_1}$. 
The remaining KK masses have been fixed as in Ref. \cite{Cheng:2002iz} 
and the SM Higgs mass is $m_h=125$\,GeV. $\Lambda R=20$ is assumed. 
The black solid line accounts for all of the dark matter (100\%) 
and the two black dotted lines show 10\% and 1\%, respectively. 
The green band shows the WMAP/Planck range, $0.117 < \Omega_{CDM}h^2 < 0.1204$.
The blue (red) solid line labelled by CDMS (XENON100) 
shows the current limit of the experiment whereas the dashed and dotted lines 
represent projected limits of future experiments. 
In the case of $\gamma_1$ LKP, a ton-scale experiment will rule out 
most of the parameter space while there is little parameter space left in the case of $Z_1$ LKP. 
The yellow region in the case of $\gamma_1$ LKP shows parameter space 
that could be covered by the collider search in the $4\ell+\met$ channel 
at the LHC with a luminosity of 100 fb$^{-1}$ \cite{Cheng:2002ab}. 
(Figures taken from \cite{Arrenberg:2013paa}.)
}
\label{fig:SI_Delta_Neutron_B_Z}
\end{figure}

In the so defined $(m_{LKP},\Delta_{q_1})$ parameter plane, in Fig.~\ref{fig:SI_Delta_Neutron_B_Z} we
superimpose the limit on the spin-independent elastic scattering cross section, 
the limit on the relic abundance and the LHC reach in the four 
leptons plus missing energy ($4\ell + \met$)
channel which has been studied in~\cite{Cheng:2002ab} at the 14 TeV 
(see Ref. \cite{Belyaev:2012ai} for 7+8 TeV). This signature
results from the pair production (direct or indirect) of $SU(2)_W$-doublet 
KK quarks, which subsequently decay to $Z_1$'s and jets. The leptons (electrons or muons)
arise from the $Z_1\to \ell^+\ell^-\gamma_1$ decay, whose branching fraction 
is approximately $1/3$~\cite{Cheng:2002ab}.
Requiring a 5$\sigma$ excess at a luminosity of 100 fb$^{-1}$, 
the LHC reach extends up to $R^{-1} \approx m_{\gamma_1} \sim 1.5$ TeV, 
which is shown as the right-most boundary of the (yellow) shaded region
in Fig.~\ref{fig:SI_Delta_Neutron_B_Z}a. The slope of that boundary is due to
the fact that as $\Delta_{q_1}$ increases, so do the KK quark masses, and their 
production cross sections are correspondingly getting suppressed, diminishing
the reach. We account for the loss in cross section according to the
results from Ref.~\cite{Datta:2005zs}, assuming also that, as expected, the 
level-2 KK particles are about two times heavier than those at level 1.
Points which are well inside the (yellow) shaded region, of course, 
would be discovered much earlier at the LHC. Notice, however, that the LHC reach 
in this channel completely disappears for $\Delta_{q_1}$ less than about 8\%.
This is where the KK quarks become lighter than the $Z_1$ (recall that 
in Fig.~\ref{fig:SI_Delta_Neutron_B_Z}a $m_{Z_1}$ is fixed according to
the MUED spectrum) and the $q_1\to Z_1$ decays are turned off. 
Instead, the KK quarks all decay directly to the $\gamma_1$ LKP 
and (relatively soft) jets, presenting a monumental challenge for an LHC discovery.
So far there have been no studies of the collider phenomenology of a
$Z_1$ LKP scenario, but it appears to be extremely challenging, especially if the 
KK quarks are light and decay directly to the LKP. This is why
there is no LHC reach shown in Fig.~\ref{fig:SI_Delta_Neutron_B_Z}b.
We draw attention once again to the
lack of sensitivity at small $\Delta_{q_1}$: such small mass splittings are 
quite problematic for collider searches. 
The current LHC exclusion limit (95\% C.L. at 8 TeV) on $R^{-1}$ is about 1250 GeV for $\Lambda R=20$ \cite{Belyaev:2012ai}. 
and this is shown as the dotted (cyan) line. The horizontal line at $\Delta_{q_1} \sim 0.2$ is the average mass splitting in MUED. 
To indicate roughly the approximate boundary of the excluded region, the slanted line around 1 TeV is added, 
assuming the shape of the boundary is similar to that for the LHC14 reach. 

In Fig.~\ref{fig:SI_Delta_Neutron_B_Z} we contrast the LHC reach 
with the relic density constraints \cite{Servant:2002aq,Kong:2005hn} and with the sensitivity of 
direct detection experiments \cite{Cheng:2002ej,Servant:2002hb}. 
The green shaded region labelled by 100\% represents the 2$\sigma$ band, 
$0.117 < \Omega_{CDM}h^2 < 0.1204$ \cite{Ade:2013zuv} and the black solid line inside this 
band is the central value $\Omega_{CDM}h^2 = 0.1187$. 
The region above and to the right of this band is disfavored
since UED would then predict too much dark matter. 
The green-shaded region is where KK dark matter 
is sufficient to explain all of the dark matter in the universe, while
in the remaining region to the left of the green band 
the LKP can make up only a fraction of the dark matter in the universe.
We have indicated with the black dotted contours the 
parameter region where the LKP would contribute only 10\% and 1\%
to the total dark matter budget. Finally, the
solid (CDMS \cite{Ahmed:2009zw} in blue and XENON100 \cite{Aprile:2012nq} in red) lines show the 
current direct detection limits, while the dotted and dashed lines 
show projected sensitivities for future experiments \cite{Aprile:2012zx,Baudis:2012bc,Baudis:2010ch}
(without rescaling by the calculated relic density).

Fig.~\ref{fig:SI_Delta_Neutron_B_Z} demonstrates the complementarity between the 
three different types of probes which we are considering. 
First, the parameter space region at very large $m_{LKP}$ is inconsistent
with cosmology -- if the dark matter WIMP is too heavy, 
its relic density is too large. The exact numerical bound on the LKP mass
may vary, depending on the particle nature of the WIMP (compare 
Fig.~\ref{fig:SI_Delta_Neutron_B_Z}a to Fig.~\ref{fig:SI_Delta_Neutron_B_Z}b)  
and the presence or absence of coannihilations (compare the
$m_{LKP}$ bound at small $\Delta_{q_1}$ to the bound at large $\Delta_{q_1}$).
Nevertheless, we can see that, in general, cosmology does provide 
an upper limit on the WIMP mass. On the other hand, colliders are 
sensitive to the region of relatively large mass splittings $\Delta_{q_1}$,
while direct detection experiments are at their best at small
$\Delta_{q_1}$ {\em and} small $m_{LKP}$. 
The relevant parameter space is therefore getting squeezed from
opposite directions and is bound to be covered eventually.
This is already seen in the case of $\gamma_1$ LKP from 
Fig.~\ref{fig:SI_Delta_Neutron_B_Z}a: the future experiments 
push up the current limit almost to the WMAP/Planck band. 
In the case of $Z_1$ LKP the available parameter space 
is larger and will not be closed with the currently envisioned experiments 
alone. However, one should keep in mind that detailed LHC studies
for that scenario are still lacking.

\begin{figure}[t]
\includegraphics[width=0.485\textwidth]{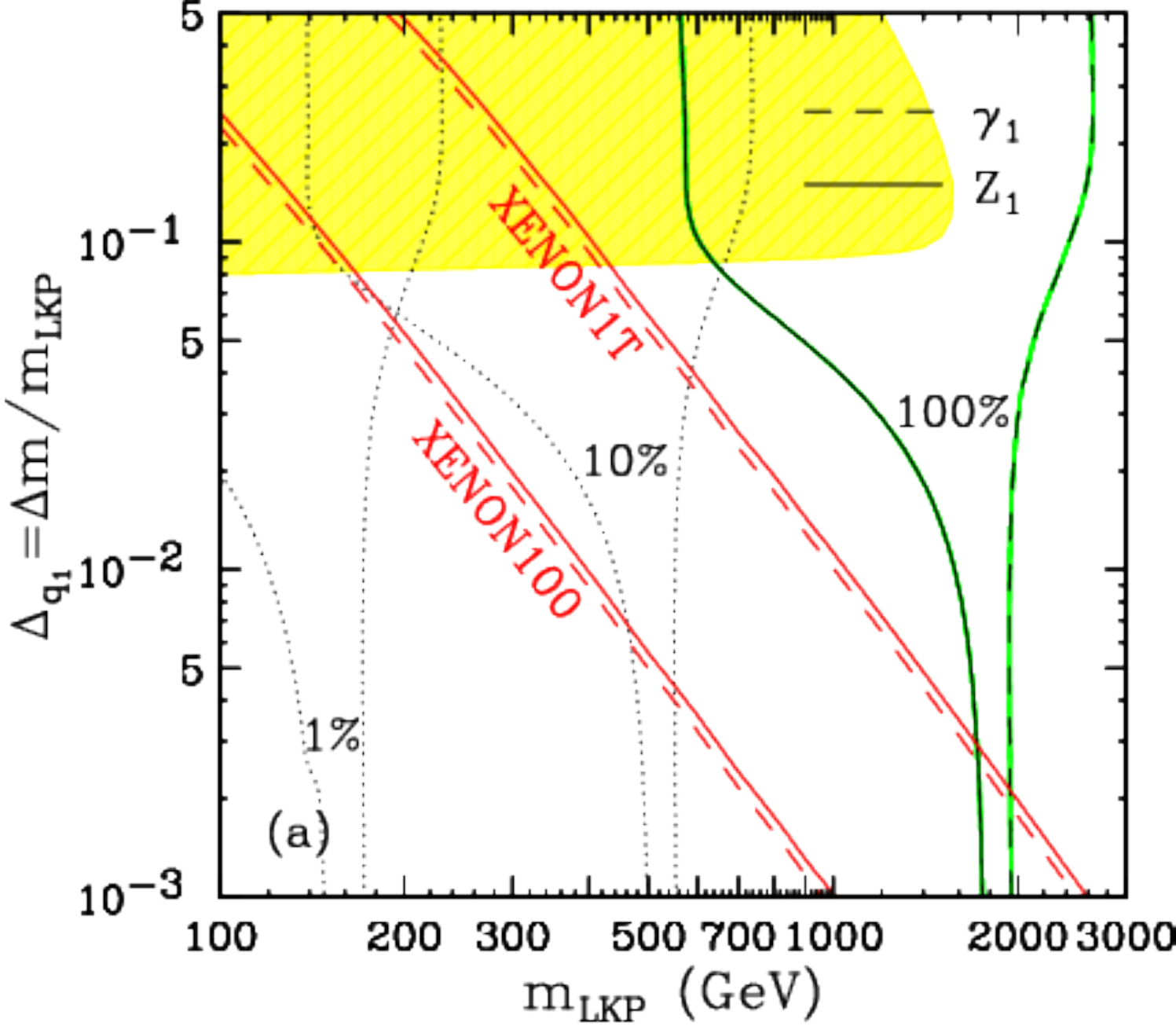}
\hspace{0.1cm}
\includegraphics[width=0.485\textwidth]{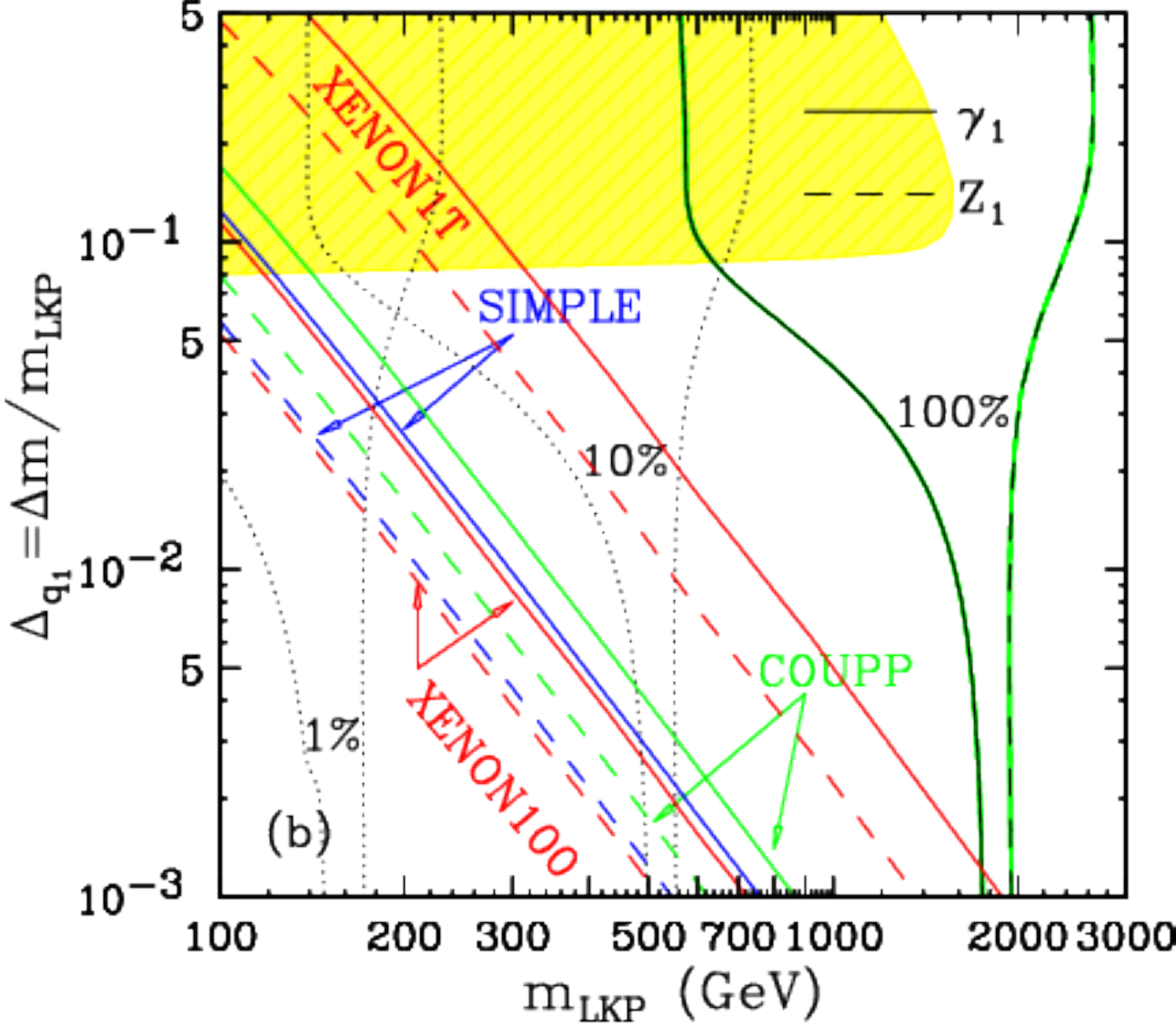}
\caption{\sl Experimental upper bounds (90\% C.L.) on the spin-dependent 
elastic scattering cross sections on (a) neutrons and (b) protons
in the $m_{LKP}$-$\Delta_{q_1}$ plane. 
The solid (dashed) curves are limits on $\gamma_1$ ($Z_1$) from each experiment. Shaded regions and dotted lines are defined in the same way as in Fig.~\ref{fig:SI_Delta_Neutron_B_Z}. The depicted LHC reach (yellow shaded region) applies only to the case of $\gamma_1$ LKP.  (Figures taken from \cite{Arrenberg:2013paa}.)
}
\label{fig:SD_Delta_B}
\end{figure}

Similarly the spin-dependent elastic scattering cross sections also exhibit an enhancement at small $\Delta_{q_1}$.
In Fig.~\ref{fig:SD_Delta_B}
we combine existing limits from three different experiments 
(XENON100 \cite{Aprile:2013doa}, SIMPLE \cite{Felizardo:2011uw} and COUPP \cite{Behnke:2012ys}) in the $(m_{LKP},\Delta_{q_1})$ plane.
Panel (a) (panel (b)) shows the constraints from the 
WIMP-neutron (WIMP-proton) SD cross sections. The rest 
of the KK spectrum has been fixed as in Fig. \ref{fig:SI_Delta_Neutron_B_Z}. 
The solid (dashed) curves are limits 
on $\gamma_1$ ($Z_1$) from each experiment. The constraints from 
LHC and WMAP on the $(m_{LKP},\Delta_{q_1})$ parameter space are the same as in 
Fig.~\ref{fig:SI_Delta_Neutron_B_Z}.

By comparing Figs.~\ref{fig:SI_Delta_Neutron_B_Z} and \ref{fig:SD_Delta_B}
we see that, as expected, the parameter space constraints for SI interactions 
are stronger than those for SD interactions. For example, in perhaps 
the most interesting range of LKP masses from 300\,GeV to 1 TeV, the SI limits 
on $\Delta_{q_1}$ in Fig.~\ref{fig:SI_Delta_Neutron_B_Z} range from $\sim 10^{-1}$ 
down to $\sim 10^{-2}$. On the other hand, the
SD bounds on $\Delta_{q_1}$ for the same range of $m_{LKP}$ are about an 
order of magnitude smaller (i.e. weaker). We also notice that the constraints 
for $\gamma_1$ LKP are stronger than for $Z_1$ LKP. This can be easily understood since 
for the same LKP mass and KK mass splitting, the $\gamma_1$ SD cross sections 
are typically larger.

Fig.~\ref{fig:SD_Delta_B} also reveals that the experiments rank 
differently with respect to their SD limits on protons and neutrons. For example,
SIMPLE and COUPP are more sensitive to the proton cross section, while
XENON100 is more sensitive to the neutron cross section.
As a result, the current best SD limit on protons comes from COUPP,
but the current best SD limit on neutrons comes from XENON100.

\subsection{Self-interacting dark matter}
\label{sec:self}

In this section, we consider dark matter that has a large cross section for scattering off of itself. We will consider simple hidden sector models where such phenomenology is realized and show that they make rather concrete predictions for direct and indirect detection experiments. This provides a simple but concrete example of complementarity between astrophysics and direct or indirect searches. The basic premise here is that dark matter particles can have can have a large cross section for elastically scattering with other dark matter particles. It is also possible that a subdominant fraction of the dark matter particles interact through dissipative collisions~\cite{Fan:2013tia}. The general scenario with collisional dark matter has been dubbed self-interacting DM (SIDM)~\cite{1992ApJ...398...43C,Spergel:1999mh}. SIDM can affect the internal structure of DM halos (density profiles and shapes) compared to collisionless DM.  In turn, astrophysical observations of structure formation, compared to N-body simulations, can probe the self-interacting nature of DM.  It is worth emphasizing that tests of self-interactions can shed light on the nature of DM {\it even if DM is completely decoupled with respect to traditional DM searches}.

In fact, there are long-standing issues on small scales that may point toward SIDM.  Dwarf galaxies are natural DM laboratories since in these galaxies DM tends to dominate baryons well inside the optical radius. Observations indicate that the central regions of well-resolved dwarf galaxies exhibit cored profiles~\cite{Moore:1994yx,Flores:1994gz}, as opposed to steeper cusp profiles found in collisionless DM-only simulations~\cite{Navarro:1996gj}.  Cored profiles have been inferred in a variety of dwarf halos, including within the Milky Way (MW)~\cite{Walker:2011zu}, other nearby dwarfs~\cite{2011AJ....141..193O} and low surface brightness galaxies \cite{2008ApJ...676..920K}.  An additional problem concerns the number of massive dwarf spheroidals in the MW.  Collisionless DM simulations have a population of subhalos in MW-like halos that are too massive to host any of the known dwarf spheriodals but whose star formation should not have been suppressed by ultraviolet feedback~\cite{BoylanKolchin:2011dk}.  While these apparent anomalies are not yet conclusive -- e.g., baryonic feedback effects may be important~\cite{Governato:2012fa} -- recent state-of-the-art SIDM N-body simulations have shown that self-interactions can modify the properties of dwarf halos to be in accord with observations, without spoiling the success of collisionless DM on larger scales and being consistent with halo shape and Bullet Cluster bounds~\cite{Vogelsberger:2012ku,Rocha:2012jg,Peter:2012jh,Zavala:2012us}.  

The figure of merit for DM self-interactions is cross section per unit DM mass, $\sigma/m_\chi$, where $\chi$ is the DM particle.  To have an observable effect on DM halos over cosmological timescales, the required cross section per unit mass must be\footnote{Here, $\sigma$ refers to the momentum-transfer weighted cross section averaged over a Maxwellian velocity distribution for a given halo with characteristic (most probable) velocity $v_0$ .  See Ref.~\cite{Tulin:2013teo} for further details.}
\be
{\sigma}/{m_\chi} \sim 1 \; {\rm cm^2/g} \; \approx \; 2 \; {\rm barns/GeV} \, ,
\ee
or larger.
From a particle physics perspective, this value is many orders of magnitude larger than the typical weak-scale cross section expected for a WIMP ($\sigma \! \sim\!1 \; {\rm picobarn}$).  Evidence for self-interactions would therefore point toward a new dark mediator particle $\phi$ that is much lighter than the weak scale.  Such light mediators have been invoked within a variety of other DM contexts as well, including explaining various indirect detection anomalies; see e.g.~\cite{Feng:2008ya,Pospelov:2008jd, ArkaniHamed:2008qn}.

As one example, DM self-interactions can arise if DM is coupled to a massive dark photon $\phi$ from a hidden $U(1)^\prime$ gauge symmetry~\cite{Feng:2009mn,Ackerman:2008gi,Feng:2009hw,Buckley:2009in,Loeb:2010gj,Tulin:2012wi,Tulin:2013teo}.  Other examples where dark matter self-interactions arise include mirror dark matter \cite{Mohapatra:2001sx,Foot:2004wz,Foot:2012ai} and atomic dark matter \cite{Kaplan:2011yj,CyrRacine:2012fz}, both appearing in the framework of hidden sector dark matter. The non-relativistic self-scattering mediated by a dark photon can be described by a Yukawa potential, 
\begin{eqnarray} \label{yukawapot}
V(r)=\pm\frac{\ax}{r} \, e^{-\mphi r}, 
\end{eqnarray} 
where $\alpha_\chi$ is the ``dark fine structure constant.''  For symmetric DM (both $\chi,\bar \chi$ are present today) scattering can be repulsive ($+$) or attractive ($-$), while for asymmetric DM (only $\chi$ is present today) scattering is purely repulsive.  Given the potential in Eq.~\eqref{yukawapot}, the cross section $\sigma$ can be computed using standard methods from quantum mechanics as a function of the three parameters $(m_\chi, m_\phi, \alpha_\chi)$ and the relative velocity $v$~\cite{Tulin:2013teo}.  

Different size DM halos have different characteristic velocities, giving complementary information about $\sigma(v)$.  
Similar to Rutherford scattering, DM self-scattering through a light mediator is typically suppressed at large velocities compared to smaller velocities.  Therefore, it is natural for DM to be self-interacting in smaller dwarf halos, while appearing to be collisionless in larger halos.  For example, the Bullet Cluster is often quoted as an example of an observation that categorically rules out self-interactions in the dark sector.  This is not true since the relative velocity in the Bullet Cluster system ($v \approx 3000\; {\rm km/s}$) is much larger than in dwarf halos ($30 \; {\rm km/s}$).  As we show below, this constraint, while important, eliminates only a small region of SIDM parameter space.

Aside from self-interactions, the mediator $\phi$ can also set the DM relic density in the early Universe through $\chi\bar{\chi}\rightarrow\phi\phi$ annihilation.  For symmetric DM, 
the required annihilation cross section is $\left<\sigma v\right>_{\rm ann} \approx 5\times10^{-26}~{\rm cm^3/s}$, which fixes 
$\ax\approx 4\times10^{-5}(\mx/{\rm GeV})$.   For asymmetric DM, although the relic density is determined by a primordial asymmetry, $\langle \sigma v \rangle_{\rm ann}$ has to be {\it larger} than in the symmetric case, implying $\ax\gtrsim 4\times10^{-5}(\mx/{\rm GeV})$.

\begin{figure}
\includegraphics[scale=0.63]{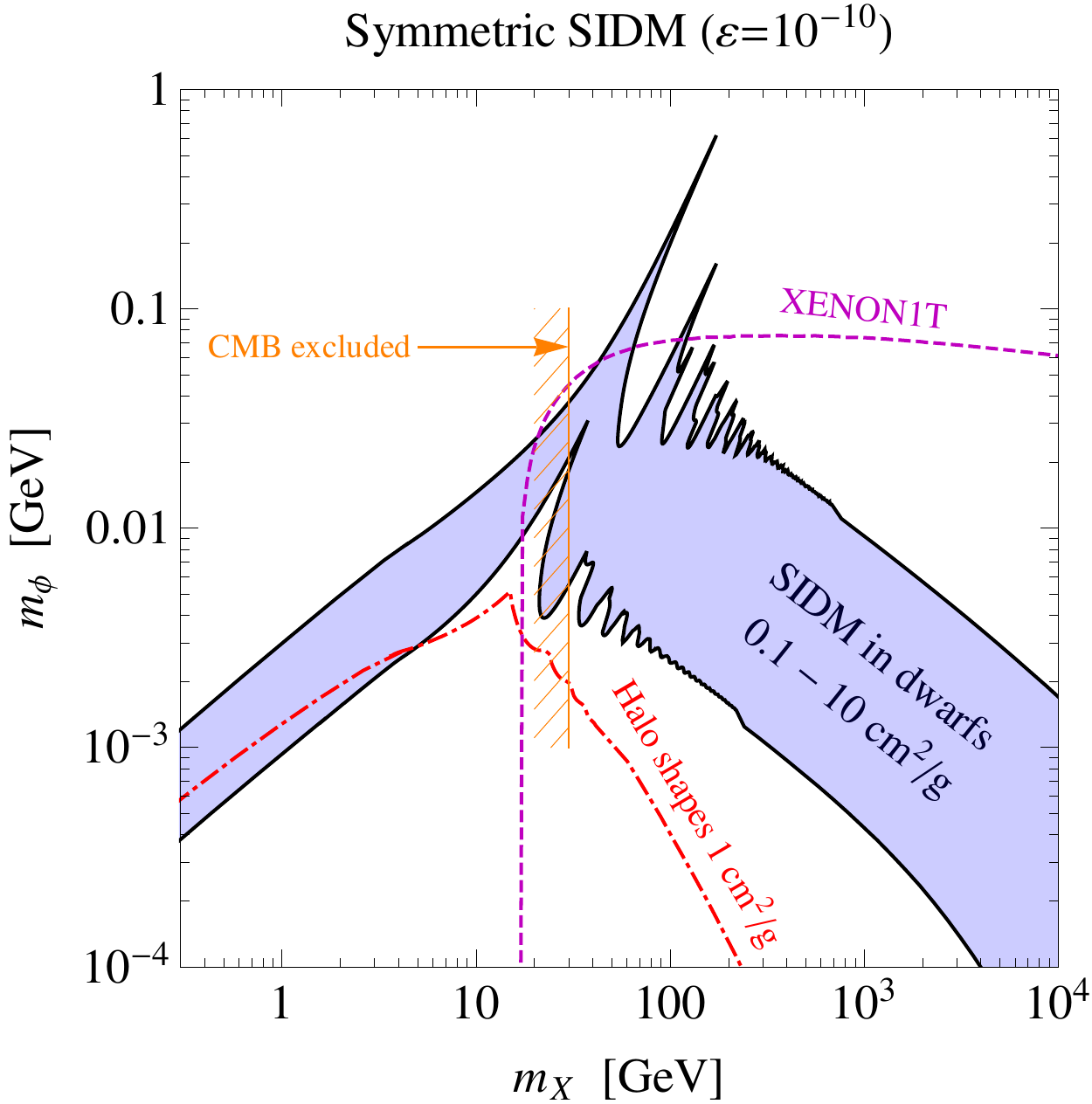}
\includegraphics[scale=0.63]{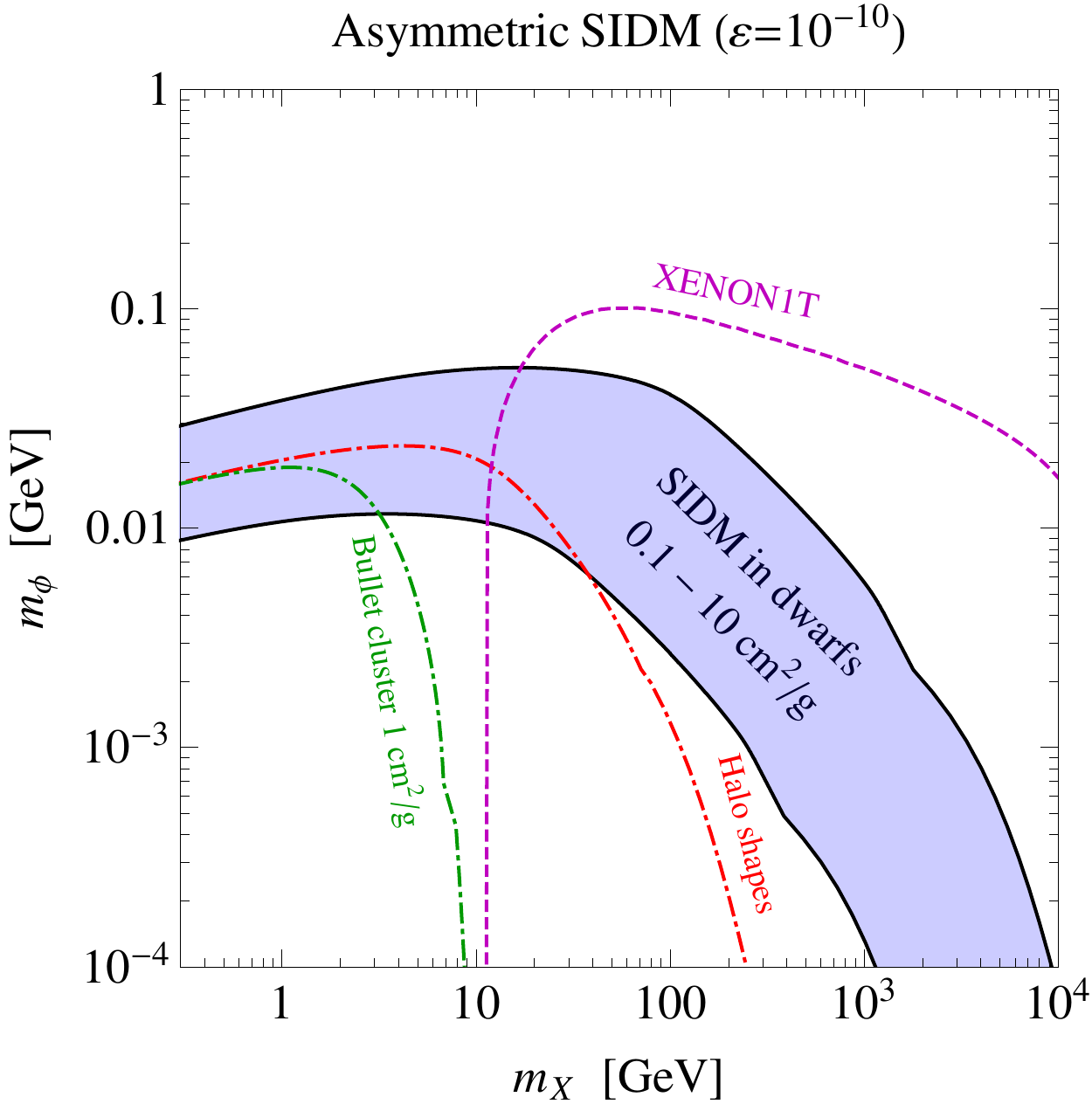}
\caption{Parameter space for SIDM $\chi$ with a vector mediator $\phi$, as a function of their masses $m_\chi,m_\phi$, for symmetric DM with $\alpha_\chi$ fixed by relic density (left) and asymmetric DM with $\alpha_\chi = 10^{-2}$ (right). Shaded region indicates the region where DM self-interactions would lower densities in the central parts of dwarf scales consistent with observations. The upper (lower) boundary corresponds to $\langle \sigma_T\rangle/m_\chi=0.1~{\rm cm}^2/{\rm g}$ ($10~{\rm cm}^2/{\rm g}$). Dot-dashed curves show halo shape constraints on group scales ($\sigma/m_\chi < 1 \; {\rm cm^2/g}$) and the Bullet Cluster constraint ($\sigma/m_\chi < 1 \; {\rm cm^2/g}$).  Dashed lines show direct detection sensitivity for XENON1T if $\phi$ has kinetic mixing with the photon with $\epsilon=10^{-10}$. The vertical hatched boundary shows exclusion from CMB if $\phi \to e^+ e^-$. See text for details. }
\label{paramspace}
\end{figure}

\begin{figure}
\includegraphics[scale=0.63]{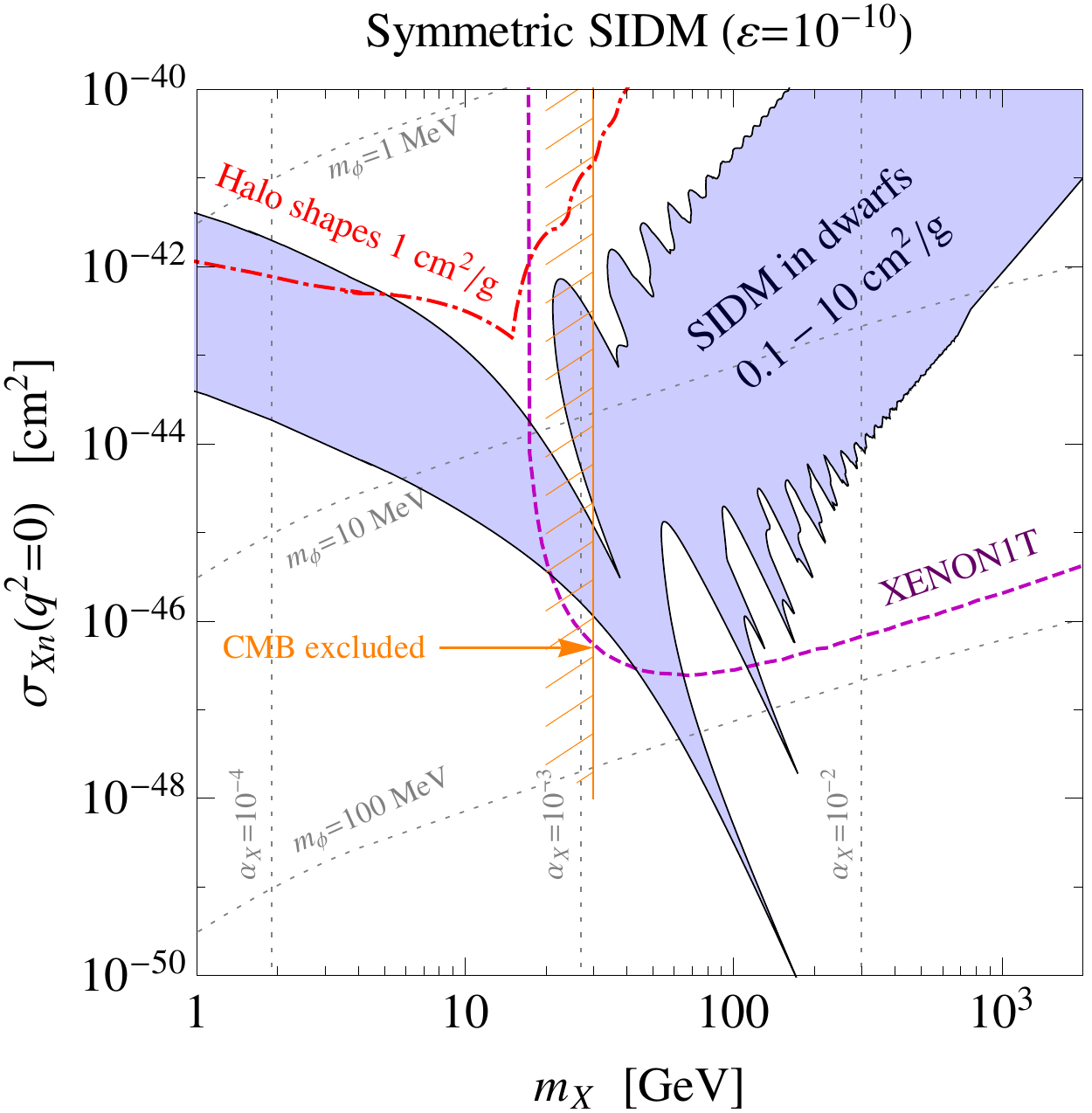} \includegraphics[scale=0.63]{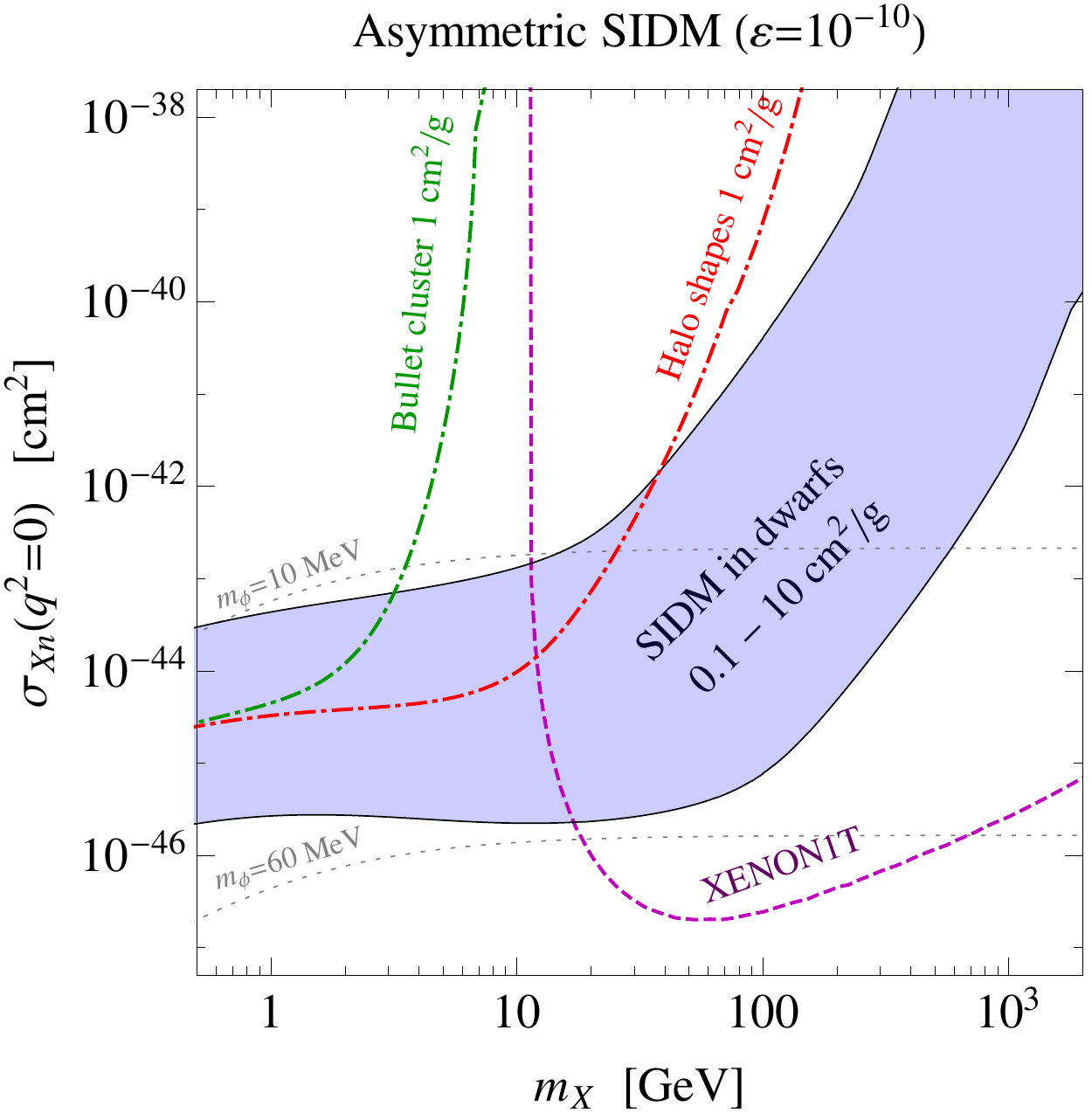}
\caption{Prospects for direct detection of self-interacting DM that couples to quarks via gauge kinetic mixing with $\epsilon=10^{-10}$.  Left figure is for symmetric DM, with $\alpha_X$ fixed by relic density constraints; right figure is asymmetric DM, with $\alpha_X = 10^{-2}$.  Shaded region indicates the region where DM self-interactions would lower densities in the central parts of dwarf scales consistent with observations. The upper (lower) boundary corresponds to $\langle \sigma_T\rangle/m_\chi=10~{\rm cm}^2/{\rm g}$ ($0.1~{\rm cm}^2/{\rm g}$). Direct detection sensitivity from future XENON1T experiments shown by dashed curves. Astrophysical limits from halo shapes and the Bullet Cluster shown by dot-dashed lines.  The range of $m_\phi, \alpha_X$ values are shown by dotted lines. The vertical hatched boundary shows exclusion from CMB if $\phi\rightarrow e^+e^-$ (left). These figures are taken from Ref.~\cite{Kaplinghat:2013kqa}.}
\label{directdet}
\end{figure}

Fig.~\ref{paramspace} shows the parameter space for this SIDM model as a function of $m_\chi$ and $m_\phi$.  The left panel corresponds to symmetric DM, where $\alpha_\chi$ is fixed by relic density, while the right panel corresponds to asymmetric DM with $\alpha_\chi = 10^{-2}$.  The shaded regions show where SIDM can explain halo anomalies on dwarf scales, with a generous range of cross section $0.1 \lesssim \sigma/m_\chi \lesssim 10\; {\rm cm^2/s}$ and taking a characteristic velocity $v_0 = 30 \; {\rm km/s}$.  The contours labeled ``SIDM'' show where $\sigma/m_\chi = 0.1, 1, 10 \; {\rm cm/s}$ on dwarf scales.  To implement the Bullet Cluster constraint, we require $\sigma/m_\chi \lesssim 1 \; {\rm cm^2/g}$ for a relative velocity $v \approx 3000\; {\rm km/s}$~\cite{Randall:2007ph}, shown by the green dot-dashed contour.  Other constraints arise from the ellipticity of DM halos of galaxies; we require $\sigma/m_\chi \lesssim 1 \; {\rm cm^2/s}$ for halos of characteristic velocity $v_0 \approx 300 \; {\rm km/s}$~\cite{Peter:2012jh}, shown by the red dot-dashed contour (``Halo shapes'').  From these bounds, the low ($m_\chi,m_\phi$) region is excluded in Fig.~\ref{paramspace}.

The dark and visible sectors need not be completely decoupled.  For example, if there exist new states charged under both the Standard Model (SM) and $U(1)^\prime$ gauge symmetries, mixing can arise between $\phi$ and the photon or $Z$ boson.  This generates effective couplings of $\phi$ to protons and neutrons, giving rise to signals in direct detection experiments.  
In the limit of zero momentum transfer, the spin-independent (SI) $\chi$-nucleon cross section can be written as
\be
\sigma_{\chi n}^{\rm SI} = \frac{16\pi \ax \alpha_{\rm em} \epsilon_{\rm eff}^2 \mu^2_{\chi n}}{m^4_\phi}  \approx 10^{-24} \; {\rm cm}^2 \times \epsilon_{\rm eff}^2 \left(\frac{30 \; {\rm MeV}}{m_\phi} \right)^4 \times \left\{ \begin{array}{cc} ({m_\chi}/{200 \; {\rm GeV}}) & {\rm symmetric \; DM} \\
({\alpha_\chi}/{10^{-2}}) & {\rm asymmetric \; DM} \end{array} \right. ,
\ee
where $\mu_{\chi n}$ is the $\chi$-nucleon reduced mass, $\alpha_{\rm em}$ is the electromagnetic fine structure constant, and $\epsilon_{\rm eff}$ is the effective $\phi$-nucleon coupling, normalized to the proton electric charge $e$.  Since SIDM prefers a very light mediator, with mass $m_\phi \sim 10 - 100$ MeV, it is clear that direct detection experiments, with current XENON100 limits approaching $\sigma_{\chi n}^{\rm SI} \sim 10^{-45}$~\cite{Aprile:2012nq}, are sensitive to very small couplings $\epsilon_{\rm eff}$.

As an example, we consider the case of kinetic mixing between $\phi$ and the photon, governed by the parameter $\epsilon$~\cite{Holdom:1985ag}.  This mixing induces a coupling of $\phi$ to SM particles carrying electric charge, so that $\phi$ decays predominantly to $e^+ e^-$ for $m_\phi$ in the $10 - 100$ MeV range prefered for SIDM.  The direct detection cross section is governed by the $\phi$-proton coupling with $\epsilon_{\rm eff} = \epsilon Z/A$, where $Z/A$ is the proton fraction of the target nucleus.  However, there are various constraints on the $\epsilon$.  Late decays of $\phi$ can inject energy to the plasma and modify standard big bang nucleosynthesis in the early Universe. 
Requiring the $\phi$ lifetime to be longer than $\sim 1$ second for leptonic decay modes, we derive a lower bound $\epsilon\gtrsim10^{-10}\sqrt{10~{\rm MeV}/\mphi}$~\cite{Lin:2011gj}. 
The upper bound from the low energy beam dump experiments is $\epsilon\lesssim10^{-7}$ for $\mphi\lesssim400~{\rm MeV}$~\cite{Bjorken:2009mm}, while the region $10^{-10} \lesssim \epsilon \lesssim 10^{-7}$ is excluded for $\mphi\lesssim100~{\rm MeV}$ by energy loss arguments in supernovae~\cite{Dent:2012mx} (although this constraint depends sensitively on assumptions about the temperature and size of the supernova core).  Regardless, for what follows, we take $\epsilon = 10^{-10}$ as a benchmark point.  

Since the mediator mass $\mphi\sim1-100$ MeV is comparable or less than the typical momentum transfer $q\sim50$ MeV in nuclear recoils, nuclear recoil interactions for SIDM are momentum-dependent and cannot be approximated by a contact interaction~\cite{Feldstein:2009tr,Chang:2009yt}. Here, we take a simplified approach by multiplying the total $q^2=0$ DM-nucleus cross section by a $q^2-$ dependent form factor: $\sigma^{\rm SI}_{\chi N}(q^2)=\sigma^{\rm SI}_{\chi N}(q^2=0)f(q^2)$, with $f(q^2)=\mphi^4/(\mphi^2+q^2)^2$. We take a fixed value $q=50$ MeV for Xenon and assume that the cross section limits quoted in the XENON experiment apply to $\sigma^{\rm SI}_{\chi N}(q^2)$ directly. We have checked that our simple approximation can reproduce the XENON100 reanalysis in~\cite{Fornengo:2011sz}.  It is interesting to note that the current XENON100~\cite{Aprile:2012nq} limits are not sensitive to our benchmark SIDM model with $\epsilon=10^{-10}$ because of the suppression from $f(q^2)$. Future direct detection experiments, such as LUX~\cite{Akerib:2012ys}, SuperCDMS~\cite{Brink:2012zza}, and XENON1T will offer great sensitivity to directly detect SIDM. In Fig.~\ref{paramspace}, we show how direct detection sensitivities from XENON1T~\cite{Aprile:2012zx} map onto SIDM parameter space for $\epsilon = 10^{-10}$ and $Z/A \approx 0.4$ (purple dashed contours).

For symmetric DM, residual annihilation can lead to additional reionization around the recombination epoch via $\chi \bar \chi \to \phi \phi \to e^+ e^- e^+ e^-$, which is constrained by CMB observations~\cite{Galli:2009zc,Slatyer:2009yq}. For ${\rm BR}(\phi \to e^+ e^-)=1$, symmetric SIDM is excluded for $\mx$ below $\sim30~{\rm GeV}$~\cite{Lopez-Honorez:2013cua}, as indicated in Fig.~\ref{paramspace} (left) with the horizontal orange line.  For asymmetric DM, this constraint does not apply.  (We also note that this bound is weakened if $\phi$ decays to neutrinos, which occurs if $\phi$ mixes with the $Z$ boson.)  

For symmetric dark matter there are further connections to indirect searches in the context of models with kinetic mixing. Annihilation to electrons and positrons locally will be constrained by the AMS-02 data~\cite{Aguilar:2013qda}, while Fermi Large Area Telescope will be sensitive to annihilation in the Galactic Center. %We do not explore this further here but refer to a paper in preparation \cite{KLY}.

In Fig.~\ref{directdet}, we illustrate the complementarity between astrophysical probes and direct detection in constraining SIDM.  Fixing $\epsilon = 10^{-10}$, we show the SIDM prediction for SI scattering in direct detection experiments for both symmetric DM (left) and asymmetric DM (right).  As in Fig.~\ref{paramspace}, the shaded band shows the prefered parameter region for solving dwarf-scale anomalies, while the red and green contours denote limits from halo shape observations and the Bullet Cluster, respectively.  The purple dashed lines show the projected XENON1T bounds~\cite{Aprile:2012zx}. The dotted gray lines denote contours of constant $\alpha_\chi$ and $m_\phi$.  This figure clearly demonstrates that astrophysical observations and direct detection experiments complement each other in the search for SIDM candidates. 

\section{Post-Discovery Complementarity}
\label{sec:postdiscovery}

As important as a broad program of complementary searches is to
establishing a compelling signal for dark matter, it becomes even more
important after a signal has been reported for several reasons.

First, as is well known, many tentative dark matter signals have
already been reported.  The potential identification of a quarter of
the Universe will require extraordinary proof in the form of
verification by other experiments.

Second, each search strategy has its limitations.  For example, as
noted in \secref{basic}, the discovery of a dark matter signal at
particle colliders only establishes the production of a particle with
lifetime greater than about 100 ns.  The assumption that this particle
contributes to dark matter requires an extrapolation in lifetime of 24
orders of magnitude! It is only by corroborating a particle collider
discovery through another method that one can claim that the collider
discovery is relevant for cosmology.

Last, the discovery of dark matter will usher in a rich and
decades-long program of dark matter studies. Consider the following
scenario: The LHC sees a missing energy signal, and precision
measurements find evidence that it is due to a 60 GeV neutralino.
This result is confirmed by direct search experiments, which discover
a signal consistent with this mass.  However, further LHC and ILC
studies constrain the neutralino's predicted thermal relic density to
be half of $\Omega_{\text{DM}}$, implying that it is not a thermal
relic, or that it makes up only half of the dark matter.  The puzzle
is resolved when axion detectors discover a signal, which is
consistent with axions making up the rest of the dark matter, and
progress in astrophysical theory, simulations, and observations leads to
a consistent picture with dark matter composed entirely of CDM.  The combined
data establish a new standard cosmology in which dark matter is
composed of equal parts neutralinos and axions, and extend our
understanding of the early Universe back to neutralino freezeout, just
1 ns after the Big Bang.  Direct and indirect detection rates are then
used to constrain the local dark matter density, halo profiles, and
substructure, establishing the new fields of neutralino and axion
astronomy.

This two-component scenario is more complicated than assumed in many
dark matter studies, but it is still relatively simple --- as is often
noted, the visible Universe has many components, and there is no
reason that the dark Universe should be any simpler.  As simple as
this scenario is, however, it illustrates the point that, even for
dark matter candidates that we have studied and understood, the
information provided by several approaches will be essential to
understanding the particle nature of dark matter and its role in
astrophysics and cosmology. A balanced program with components in each
of the four approaches is required to cover the many well-motivated
dark matter possibilities, and their interplay will likely be
essential to realize the full potential of upcoming discoveries.

\section{Conclusions}
\label{sec:conclusions}

The problem of identifying dark matter is central to the fields of
particle physics and astrophysics, and has become a leading problem in
all of basic science.  In the coming decade, the field of dark matter
will be transformed, with a perfect storm of experimental and
technological progress set to put the most promising ideas to the
test.

Dark matter searches rely on four approaches or pillars: direct
detection, indirect detection, particle colliders, and astrophysical
probes.  In this Report, we have described the complementary relation
of these approaches to each other.  This complementarity may be seen
on several levels.  First, these approaches are qualitatively
complementary: they differ in essential characteristics, and they rely
on different dark matter properties to see a signal.  A complementary
set of approaches is required to be sensitive to the dark matter
possibilities that are currently both viable and well-motivated.  The
approaches are also quantitatively complementary: within a given class
of dark matter possibilities, these approaches are sensitive to
different dark matter interactions and mass ranges.

Last, the discovery of a compelling dark matter signal is only the
beginning.  Complementary experiments are required to verify the
initial discovery, to determine whether the particle makes up all of
dark matter or only a portion, and to identify its essential
properties, such as its interactions, spin, and mass, and to determine
its role in forming the large scale structures of the Universe that we
see today.  A balanced dark matter program is required to carry out
this research program to discover and study dark matter and to
transform our understanding of the Universe on both the smallest and
largest length scales.

\vspace{1cm}
{\bf Acknowledgements}: MK would like to thank David Weinberg and Anze Slosar for input on section \ref{sec:astrophysics}.

%\newpage

\bibliography{bibdmcomp}{}

\end{document}